\RequirePackage{fix-cm}
\documentclass{svjour3}                     
\smartqed  
\usepackage{graphicx}
\usepackage{graphicx,epstopdf}
\usepackage{dcolumn}
\usepackage{bm}

\begin{document}

\title{Cosmic Time Transformations in Cosmological Relativity}

\titlerunning{Cosmic time transformations}

\author{Firmin J. Oliveira \\
P. O. Box 10882 \\ Hilo, Hawaii, U.S.A.  96721-5882   \\ Email address: firmjay@hotmail.com}

\authorrunning{Firmin J. Oliveira}

\date{Received:  / Accepted:}


\maketitle


\begin{abstract}
The relativity of cosmic time is developed within the framework of Cosmological Relativity in five 
dimensions of space, time and velocity.    A general linearized metric element is defined to have 
the form $ds^2 = (1+\phi) c^2 dt^2  - dr^2 + (1+\psi) \tau^2 dv^2$, where the coordinates 
are time $t$, radial distance $r=\sqrt{x^2 + y^2 + z^2}$ for spatials $x$, $y$ and $z$,
and velocity $v$, with $c$ the speed of light in vacuum and $\tau$ the Hubble-Carmeli time
constant.  The metric is accurate to first order in $t/\tau$ and $v/c$. The fields $\phi$ and $\psi$ 
are general functions of the coordinates.
By showing that $\phi = \psi$, a metric of the form $ds^2 = c^2 dt^2  - dr^2 + \tau^2 dv^2$
is obtained from the general metric, implying that the universe is flat.   For cosmological redshift $z$,
the luminosity distance relation $D_L (z,t) = r (1 + z) / \sqrt{1 - t^2 / \tau^2}$ is used to fit combined 
distance moduli from Type Ia Supernovae up to $z < 1.5$ and Gamma-Ray Bursts up to $z < 7$, 
from which a value of $\Omega_M = 0.800 \pm 0.080$ is obtained for the matter density parameter at 
the present epoch.  Assuming a baryon density of $\Omega_B = 0.038 \pm 0.004$, a rest mass energy  of 
$( 9.79  \pm  0.47 ) \, {\rm GeV}$ is predicted for the anti-baryonic $\bar{Y}$ and the $\Phi^{*}$ particles 
which decay from a hypothetical $\bar{X}_1$ particle.  The cosmic aging function 
$g_1(z,t)= ( 1 + z) ( 1 - t^2 / \tau^2 )$ makes good fits to light curve data from two reports of Type 1a 
supernovae and in fitting to simulated quasar like light curve power spectra separated by redshift 
$\Delta{z} \approx 1$.   We determine the multipole of the first acoustic peak of the Cosmic Microwave 
Background radiation anisotropy to be $l \approx 224 \pm 5$ and a sound horizon 
of $\theta_{sh0} \approx  (0.805 \pm 0.020 ) {}^{\circ}$ on today's sky.
\end{abstract}



\section{Introduction}
  It has recently been reported\cite{hawkins-1} on the apparent null effect of cosmic time dilation
  upon light curve power spectra measurements of some 800 low and high redshift quasars (QSO)
  monitored for 28 years. This appears to contradict two Supernovae Type Ia (SNe-Ia) light curve evolution
  studies\cite{goldhaber-1,blondin-1} which show the effect of broadening of the power
  spectra time series consistent with a cosmic time dilation of $(1 + z)$ for redshift $z$.
  In this paper we show that both the QSO and SNe-Ia results are compatible if account is made for the
  relativity of cosmic time as developed in the theory of Cosmological Special Relativity (CSR)\cite{carmeli-0}.
  We also apply these concepts to fitting the combination of high redshift SNe-Ia distance data\cite{SCPUnion-1}
  and Gamma-Ray Bursts (GRB) data\cite{schaefer-1}

  The Cosmological Relativity of Carmeli\cite{carmeli-2}, the general and special theories, is a five dimensional
  brane world model based on time $t$, space $x, y, z$ and velocity $v$.  It makes only one new assumption,
  that of the maximum value of cosmic time $\tau$ which is the Hubble-Carmeli time constant.  In the same way
  that the constant speed of light $c$ constrains the observations in space and time,
  so too the constant of cosmic time $\tau$ constrains the observations of space and velocity.
  The familiar Lorentz transformations of Einstein Special Relativity (SR) between observers moving at constant
   relative velocity $v$ carry over into Cosmological Special Relativity (CSR) between observers separated by 
   relative cosmic time $t$.  And just as in SR there is the special way that velocities add together which reduces 
   to the Galilean form $v_1 + v_2$ at  low velocities with respect to the speed of light, in CSR cosmic times add 
   in an analogous way which has the form $t_1 + t_2$ for low values with respect to cosmic time $\tau$ 
   but is modified for larger cosmic times.  

   In this paper we derive only the minimum CSR framework we require for the development of cosmic  time 
   transformation effects and the luminosity distance relation.    We show that for weak fields $\phi$ and $\psi$
   (to be defined below) the universe has a flat, Euclidean geometry.  And, due to the additive properties of cosmic
   time in CSR, this gives us a unique form for the luminosity distance relation.  We model cosmic time aging 
   effects in light curve data from SNe-Ia. 
   In comparison to the standard  Friedmann-Lema\^{i}tre-Robertson-Walker (FLRW) model, the CGR luminosity 
   distance relation performs quite well in fits to SNe-Ia and GRB  distance data,  although it requires more dark 
   matter in the mass density.  As a consequence, larger rest masses are predicted for the hypothetical $\bar{X}_1$ 
   particle\cite{davoudiasl-1} decay products $\bar{Y}$ and $\Phi^{*}$.  We also derive a relation for the first acoustic 
   peak of the Cosmic Microwave Background (CMB) anisotropy.

\section{The universe}

The five dimensional Cosmological General Relativity of Carmeli\cite{carmeli-2} is approximated by the
linearized metric element,
\begin{equation}
ds^2 = (1+\phi) c^2 dt^2  - dr^2 + (1+\psi) \tau^2 dv^2,    \label{eq:5Dmetric}
\end{equation}
where
\begin{equation}
dr^2 = dx^2 + dy^2 + dz^2.    \label{eq:dr2}
\end{equation}
The coordinates are time $t$, spatials $x$, $y$ and $z$ and velocity $v$, with 
$c$ the speed of light in vacuum and $\tau$ the Hubble-Carmeli time constant. The linearized model is accurate to 
first order in $v/c \ll 1$ and $t/\tau \ll 1$.
The parameter $h = 1 / \tau$ is the Hubble constant at zero distance and no gravity and $h \approx H_0$ where $H_0$
is the Hubble constant. The fields $\phi$ and $\psi$ are general functions of the coordinates. The CGR standard value for
$h$ is  (to three digits)\cite{oliveira-0}, 
\begin{equation}
  h  =  72.2 \pm 0.84 \, {\rm km \, s^{-1} \, Mpc^{-1}}.    \label{eq:h-value} 
\end{equation}
This gives
\begin{equation}
  \tau = \left( 4.28 \pm 0.15 \right)   \times 10^{17} {\rm s}  = 13.6 \pm 0.48 \, {\rm Gyr}.  \label{eq:tau-value}
\end{equation}
In Cosmological Relativity the time coordinate is measured backwards from the present at $t=0$ to the big bang at $t = \tau$.
Just as the speed of light $c$ is the maximum observable speed, the time $\tau$ is the maximum observable cosmic time.
The gravitational fields are specified by the functions $\phi$  and  $\psi$. The expansion of the universe occurs when 
\begin{equation}
ds = 0,  \label{eq:ds=0}
\end{equation}
 and it is observed at a specific cosmic time $t$, so that 
\begin{equation}
dt=0.  \label{eq:dt=0}
\end{equation}
Applying (\ref{eq:ds=0}) and (\ref{eq:dt=0}) to (\ref{eq:5Dmetric}) yields for the expanding universe 
\begin{equation}
-dr^2 + (1+\psi) \tau^2 dv^2  =  0, \label{eq:expansion}
\end{equation}
which reduces to
\begin{equation}
\left( \frac{dr} {\tau dv} \right)^2  = 1 + \psi.   \label{eq:dr-by-dv}
\end{equation}
The metric  (\ref{eq:5Dmetric}) defines the Einstein field equations in five dimensions 
\cite[Sect.~7.3]{carmeli-2}
\begin{equation}
R^{\nu}_{\mu}  - \frac{1} {2} \delta^{\nu}_{\mu} R = \kappa T^{\nu}_{\mu},  \label{eq:einfield-eq}
\end{equation}
where $R^{\nu}_{\mu}$ is the mixed Ricci tensor\cite[Appendix A)]{carmeli-2}, $R$ is the Ricci scalar and 
$T^{\nu}_{\mu} = \rho_{eff} u_{\mu} u^{\nu}$ is the mixed energy-momentum tensor, where $\rho_{eff}$
is the effective mass density and $u_{\mu} = u^{\mu} =(1,0,0,0,1)$ is the velocity vector. 
The indices $\nu, \mu  = 0, 1, 2, 3, 4$ for the five dimensions of $x^0 = c t, x^1 = x, x^2 = y, x^3 = z, x^4 = \tau v$.  
The Kronecker delta $\delta^{\nu}_{\mu} = 1$ for $\mu = \nu$,
and $\delta^{\nu}_{\mu} = 0$ for $\mu \ne \nu$.  The Carmeli gravitation constant $\kappa = 8 \pi G / c^2 \tau^2$ 
where $G$ is Newton's gravitation constant. The $0,0$ component of (\ref{eq:einfield-eq}) gives us the
equation
\begin{equation}
R^0_0  - \frac{1} {2}  R  = \kappa \tau^2 \rho_{eff},  \label{eq:einfield-eq-00}
\end{equation}
where
\begin{eqnarray}
R^0_0 - \frac{1} {2} R &=& \frac{1} {2} \left( \nabla^2 \phi - \phi_{,44}  - \psi_{,00} \right) 
    - \frac{1} {2} \left(  \nabla^2 \phi +  \nabla^2 \psi   - \phi_{,44}  - \psi_{,00} \right)  \label{eq:R00-halfR-1}, \\
\nonumber \\
                                   &=& -\frac{1} {2} \nabla^2 \psi,  \label{eq:R00-halfR-2}
\end{eqnarray}
where $\psi_{,00} = \partial^2{\psi} / \partial{t^2}$ and $\phi_{,44} = \partial^2{\phi} / \partial{v^2}$.
From (\ref{eq:einfield-eq-00}) and (\ref{eq:R00-halfR-2}) we get the Poisson equation for cosmology,
in the space-velocity domain,
\begin{equation}
\nabla^2 \psi  = -2 \kappa \tau^2 \rho_{eff}.  \label{eq:basic-diff-eqn}
\end{equation}
The effective mass density is defined by  
\begin{equation}
\rho_{eff} = \rho - \rho_c,  \label{eq:rho_eff-def}
\end{equation}
where $\rho$ is the mass density and $\rho_c$ is the critical mass density.  Under the assumption that the 
mass is uniformly distributed, the mass density $\rho$ is independent of the spatial coordinate $r$, 
but can depend on time $t$ and velocity $v$. The critical mass density $\rho_c$ is a constant defined by
\begin{equation}
 \rho_c = 3 / 8 \pi G \tau^2. \label{equation:critical-mass}
\end{equation}
It is useful to express the effective mass density as a parameter in terms of the critical mass density.  Dividing by 
$\rho_c$ we have
\begin{equation}
\Omega_{eff} =  \frac{\rho_{eff}} { \rho_c }   =  \Omega  - 1,   \label{eq:Omega_eff}
\end{equation}
where $\Omega = \rho / \rho_c$ is the mass density parameter.  For a spatially uniform mass distribution,
$\rho$ is independent of $r$, the solution to (\ref{eq:basic-diff-eqn}) then takes the form 
(Ref. \cite[Sect.7.3.2]{carmeli-2}),
\begin{equation}
\psi =   -\frac{\Omega_{eff} \, r^2} { c^2 \tau^2 }  - \frac{2 G M} {c^2 r },  \label{eq:psi-solution}
\end{equation}
where $M$ is an optional point mass centered at the origin of $r$.  Assuming no central mass, we set $M=0$. 
Then putting the expression for $\psi$ from (\ref{eq:psi-solution}) 
into (\ref{eq:dr-by-dv}) and simplifying we get
\begin{equation}
\frac{dr} {dv} = \tau \sqrt{1 - \Omega_{eff}  \, r^2  /  c^2 \tau^2  }.  \label{eq:dr-by-dv-solve}
\end{equation}

\section{General solution in space-velocity}

The integration of (\ref{eq:dr-by-dv-solve}) over $r$ and $v$ in the space-velocity domain is carried out 
at a specific time $t$.  We assume that the mass density $\rho$ is a function of cosmic time only, so that 
$\Omega_{eff}$ will be constant throughout the integration. Substitute $\Omega - 1 = \Omega_{eff}$ by 
(\ref{eq:Omega_eff}) and put (\ref{eq:dr-by-dv-solve}) into the integral form
\begin{equation}
\int^r_0{ \frac{dr'} {\sqrt{1 + \left( 1 - \Omega \right) r'^2  /  c^2 \tau^2  } } } =   \int^v_0{\tau dv'}. 
       \label{eq:integral-dr}
\end{equation}
Integrating (\ref{eq:integral-dr}) and solving for $r$ in terms of $v$ we obtain the general solutions
\begin{eqnarray}
r  &=&  \frac{ c \tau } { \sqrt{ 1 - \Omega } }  \, {\rm sinh}\left( \frac{v} {c}  \sqrt{ 1 - \Omega } \right), 
                   \;  \; {\rm for} \; \Omega < 1,   \label{eq:r-intermsof-v-with-Omega_eff<0}  \\
\nonumber \\
r  &=&  \frac{ c \tau } { \sqrt{ \Omega - 1} }  \, {\rm sin}\left( \frac{v} {c}  \sqrt{ \Omega - 1 } \right),
                    \; \; {\rm for} \; \Omega > 1 \; {\rm and}  \label{eq:r-intermsof-v-with-Omega_eff>0} \\
\nonumber \\
r  &=&  \tau v,  \; \; {\rm for} \; \Omega = 1.  \label{eq:r-intermsof-v-with-Omega_eff=0}
\end{eqnarray}
By use of the identities ${\rm sinh}( i \, x ) =  i \, {\rm sin}( x )$ and 
${\rm  sin}( i \, x ) = i \, {\rm sinh}( x )$, where $x$ is real and $i = \sqrt{-1 \,}$, we write the general 
solution as
\begin{equation}
r  =  \frac{ c \tau } { \sqrt{ 1 - \Omega} }  \, {\rm sinh}\left( \frac{v} {c}  \sqrt{ 1 - \Omega} \right), 
                   \;  \; {\rm for} \; 0 \le \Omega.  \label{eq:r-with-any-Omega} 
\end{equation}

\section{Flat space metric}

Equation (\ref{eq:psi-solution}) gave the solution for the field $\psi$.
Now consider the solution for the field $\phi$ in (\ref{eq:5Dmetric}).
The $4,4$ component of (\ref{eq:einfield-eq}) gives us the equation\cite{carmeli-2}
\begin{equation}
R^4_4  - \frac{1} {2}  R  = \kappa \tau^2 \rho_{eff},  \label{eq:einfield-eq-44}
\end{equation}
where
\begin{eqnarray}
R^4_4 - \frac{1} {2} R &=& \frac{1} {2} \left( \nabla^2 \psi - \phi_{,44}  - \psi_{,00} \right) 
    - \frac{1} {2} \left(  \nabla^2 \phi +  \nabla^2 \psi   - \phi_{,44}  - \psi_{,00} \right)  \label{eq:R44-halfR-1}, \\
\nonumber \\
                                   &=& -\frac{1} {2} \nabla^2 \phi,  \label{eq:R44-halfR-2}
\end{eqnarray}
where $\psi_{,00} = \partial^2{\psi} / \partial{t^2}$ and $\phi_{,44} = \partial^2{\phi} / \partial{v^2}$.
Similar to the case for obtaining $\psi$, we solve (\ref{eq:R44-halfR-2}) to obtain
\begin{equation}
\phi =   -\frac{\Omega_{eff} \, r^2} { c^2 \tau^2 }  - \frac{2 G M} {c^2 r },  \label{eq:phi-solution}
\end{equation}
where $M$ is an optional point mass centered at the origin of $r$.  We see by 
(\ref{eq:psi-solution}) and (\ref{eq:phi-solution}) that in fact,
\begin{equation}
 \phi(r) = \psi(r),   \label{eq:phi=psi}
\end{equation}
 and furthermore, that the constants $c$ and $\tau$ are part of the (cosmological) first term while the constants
$c$ and $G$ are part of  the (Newtonian) second term.  Assuming no central mass ($M=0$), for the case where 
the mass density becomes equal to the critical density, $\rho \rightarrow \rho_c$, from (\ref{eq:Omega_eff}), 
the effective mass density parameter $\Omega_{eff} = \Omega - 1 \rightarrow 0$ and the universe becomes Euclidean.

On the other hand, more generally, we can derive a flatspace metric. Given (\ref{eq:phi=psi}), if we divide 
(\ref{eq:5Dmetric}) by $1+\phi = 1 + \psi$, we can write
\begin{equation}
d{\bar{s}}^2  = c^2 dt^2  - d{\bar{r}}^2 + \tau^2 dv^2,  \label{eq:new-flat-metric}
\end{equation}
where
\begin{equation}
\bar{s} = \int{\frac{ds} {\sqrt{1+\phi}} },  \label{eq:sbar}
\end{equation}
and
\begin{equation}
\bar{r} = \int{\frac{dr} {\sqrt{1+\phi}} },  \label{eq:rbar}
\end{equation}
with the condition that
\begin{equation}
-1 < \phi.  \label{eq:phi-conditional}
\end{equation}
This implies that for weak fields $\phi$ and $\psi$ we can express the cosmology of the expanding universe
in terms of a flat space Euclidean geometry, with a metric of the form $ds^2 = c^2 dt^2 - dr^2 + \tau^2 dv^2$.
In the next section we derive this flat space special theory for cosmology.

\section{The cosmological special relativistic transformation}

By (\ref{eq:new-flat-metric}), with notation $\bar{r} \rightarrow r$, we have the metric
\begin{equation}
ds^2 = c^2 dt^2 - dr^2 + \tau^2 dv^2,  \label{eq:euclidean-metric}
\end{equation}
where $r$ is given by (\ref{eq:rbar}).
The expansion of the universe occurs when $ds=0$ and observations are made at a particular instant of time $t$
so that $dt=0$.  Then for the expansion, (\ref{eq:euclidean-metric}) gives us
\begin{equation}
dr = \tau dv, \label{eq:hubble-law-differential}
\end{equation} 
which, upon integration gives the Hubble law
\begin{equation}
v =  r / \tau   = h r, \label{eq:hubble-law}
\end{equation}
where $h=1/\tau$.

In the observable universe there are two classes of objects, those that are bounded by the gravitation of their combined 
masses and those that are observed to be moving away from one another in the Hubble flow.  In other words,
if we lump all nearby neighboring galaxies into a super galaxy mass point, then the universe would consist of only
super galaxy mass points flying apart in the Hubble flow.  Cosmological Special Relativity (CSR) describes 
these super galactic objects in the universe.   However, unlike SR which can have real observers in reference frames
which move relatively at less than light speed, in CSR, all galaxies are in the Hubble flow and expand
at the Hubble rate $h$. That is, there are no objects not in the Hubble flow from which to set up a frame of
reference and compare observations.
Consequently, for CSR we define hypothetical observers in their frames which move relatively at a rate $1/t > h$. 
We derive the transformation of coordinates between these hypothetical frames.   However, the coordinates of real 
galaxies are included in the transformation.

In CSR the age of the universe is $\tau$, the Hubble-Carmeli time constant, which is assumed  to be the same for all 
{\em cosmic time relative} inertial observers and is the maximum cosmic time (just as in SR the speed of light $c$ is the 
maximum velocity and is constant for all {\em velocity relative} inertial observers.)   Assume that there are hypothetical
observers in reference frames $K$ and  $K^{'}$ separated by a fixed cosmic time  $t < \tau$.  Each frame is assumed
to be unaccelerated (inertial) with respect to cosmic time. Transformations are made between $K$ describing an object 
$O$ with ``4-vector'' coordinates $(\tau v,x,y,z)$ and $K'$ describing the same object  $O$ with 4-vector coordinates 
$(\tau v',x',y',z')$. The magnitude of each 4-vector is defined by
\begin{eqnarray}
  S^2 &=&  \tau^2 v^2 - x^2 - y^2 - z^2 =  \tau^2 v^2 - r^2,  \label{eq:distance-S} \\
\nonumber \\
  S'^2 &=&  \tau^2 v'^2 - x'^2 - y'^2 - z'^2 =  \tau^2 v'^2 - r'^2,   \label{eq:distance-S'}
\end{eqnarray}
with the invariant condition
\begin{equation}
S^2  =  S'^2,  \label{eq:S2=S'2-invar}
\end{equation}
where $r=\sqrt{x^2 + y^2 + z^2}$ and $r'=\sqrt{x'^2 + y'^2 + z'^2}$.
For the case that object $O$ is a galaxy in the expansion, the invariants $S^2 = S'^2 = 0$. We refer to $S$ 
and $S'$ as 4-vectors even though there are only 2 components, since the 3 spatial components $(x,y,z)$ are 
condensed into $r = \sqrt{x^2 + y^2 + z^2}$.

To obtain the cosmological transformation (Ref. \cite[Sect.~2.2]{carmeli-2}), analogous with the Lorentz 
transformation, assume that a linear transformation exists between the coordinates  of hypothetical 
frames $K$ and  $K'$.  In frame $K$ define the space-velocity 4-vector $S = ( \tau v, r )$ 
with magnitude
\begin{equation}
S^2 = \tau^2 v^2 - r^2  \ge 0.
       \label{eq:r_by_v_not=tau}
\end{equation}
Similarly, in frame $K'$ define the space-velocity 4-vector $S' = ( \tau v', r' )$ with magnitude
\begin{equation}
S'^2 = \tau^2 v'^2 - r'^2 \ge 0.
       \label{eq:r'_by_v'_not=tau}
\end{equation}
By the requirement that the magnitude of a 4-vector is invariant under transformations between reference frames
then  
\begin{equation}
S^2  =  S'^2.   \label{eq:S2=S'2_invariance}
\end{equation}

Clearly, by (\ref{eq:r_by_v_not=tau}) and (\ref{eq:r'_by_v'_not=tau}), the 4-vectors $S$ and  
$S'$ describe coordinates which can be either in the Hubble flow, when  $S^2 = S'^2 = 0$, or not in the Hubble flow,
when $S^2 = S'^2 >0$.  In order to obtain the transformation equations between the 4-vectors it is required that the 
reference frames $K$ and $K'$ be separated by a fixed cosmic time of $t  < \tau$.

Then, for constants $\cosh{\left( \sigma \right)} $ and  $\sinh{\left( \sigma \right)}$, where $\sigma$ is a constant 
hyperbolic angle, define $r'$ and $v'$ such that
\begin{equation} 
r'  =  r \cosh{\left( \sigma \right)} - \tau v \sinh{\left( \sigma \right)},    \label{eq:transfrm-r'-rv}
\end{equation}
and
\begin{equation}
\tau v'  =  \tau v \cosh{\left( \sigma \right)} - r \sinh{\left( \sigma \right)}.    \label{eq:transfrm-v'-rv}  
\end{equation}
To solve for the angle $\sigma$ use the  boundary condition $r' = 0$  which represents the origin of frame $K'$. 
At $r'=0$  (\ref{eq:transfrm-r'-rv}) yields,
\begin{equation}
\tanh{\left( \sigma \right)} = \frac{\sinh{\left( \sigma \right)} } {\cosh{\left( \sigma \right)} }
   = \frac{r} {\tau v}  =  \frac{t} {\tau},  \label{eq:tanhsigma-r-by-tauv}
\end{equation}
where  
\begin{equation}
t   =  \frac{r} {v}   \label{eq:t=r/v<=tau}
\end{equation}
is the fixed cosmic time separating the origins of frame $K'$ relative to $K$.  From (\ref{eq:tanhsigma-r-by-tauv}),
for
\begin{equation}
t \le \tau,  \label{eq:t-lt-tau}
\end{equation}
where $t = \tau$ is a limiting condition, we use the hyperbolic functional identities to obtain, 
\begin{eqnarray}
\cosh{\left( \sigma \right)}  &=&  \frac{ 1 } { \sqrt{ 1 - \tanh^2{\left( \sigma \right)} } } 
       = \frac{ 1 } { \sqrt{ 1 - t^2 / \tau^2 } },  \label{eq:cosh-sigma} \\
\nonumber \\
\sinh{\left( \sigma \right)}  &=&  \frac{ \tanh{\left( \sigma \right)} } { \sqrt{ 1 - \tanh^2{\left( \sigma \right)} } } 
       = \frac{ t / \tau } { \sqrt{ 1 - t^2 / \tau^2 } }.  \label{eq:sinh-sigma}
\end{eqnarray}
Substituting from (\ref{eq:cosh-sigma}) and (\ref{eq:sinh-sigma}) into (\ref{eq:transfrm-r'-rv}) and
(\ref{eq:transfrm-v'-rv}) we obtain
\begin{eqnarray}
r' &=& \frac{r - v t} {\sqrt{1 - t^2 / \tau^2} },   \label{eq:r'-trans}   \\
\nonumber \\
\tau v' &=& \frac{\tau v - r t /\tau} {\sqrt{1-t^2/\tau^2}}. \label{eq:v'-trans}
\end{eqnarray}
By inverting (\ref{eq:r'-trans}) and (\ref{eq:v'-trans}) we get the inverse
transform equations
\begin{eqnarray}
r &=& \frac{r' +  v' t} {\sqrt{1 - t^2 / \tau^2} },   \label{eq:r-trans}  \\
\nonumber \\
\tau v &=& \frac{\tau v' + r' t / \tau} {\sqrt{1 - t^2/\tau}}.   \label{eq:v-trans}
\end{eqnarray}
It can be verified that the transformations (\ref{eq:r'-trans}-\ref{eq:v-trans}) satisfy the invariance requirement
of (\ref{eq:S2=S'2_invariance}) 
that $S^2 = S'^2$  by direct substitution into (\ref{eq:distance-S}) and (\ref{eq:distance-S'}).
As was previously mentioned, for a galaxy O with 4-vector $S=(\tau v, r)$ observed by the observer in frame $K$,
where $S^2 = \tau^2 v^2  - r^2 = 0$, the transformation equations (\ref{eq:r'-trans}) and 
(\ref{eq:v'-trans}) gives for the galaxy O observed in $K'$ the 4-vector $S' = (\tau v', r')$ where, by the 
invariance of the transformation, $S'^2 = \tau^2 v'^2  - r'^2 = 0$.  Thus, Hubble coordinates are conserved.   
In the next section we will use this property to obtain the cosmological redshift relation.

\section{Cosmological redshift of light}

We wish to quantify observations of light wave phenomena in the expanding universe made by observers at 
different cosmic times.
Consider the distance $r$ to a galaxy $O$ which is in the Hubble flow, measured by the observer at the origin of $K$.  
For the observer at the origin of $K'$ the distance to the same galaxy $O$ is $r'$.  If $r = N \lambda$ and 
$r' = N {\lambda}'$,where the $\lambda$'s are the measured wavelengths of the light from the galaxy and $N$ is the 
fixed number of wavelengths, then taking the ratio of distances we get
\begin{equation}
\frac{r} {r'}  = \frac{N \lambda} {N {\lambda}'}  =  \frac{\lambda} {{\lambda}'}  = 1 + z,  \label{eq:cosmo-redshift}
\end{equation} 
where $z$ is the cosmological redshift of the light due to the expansion of space during the cosmic time $t$ 
between frames $K'$ and $K$ . Substituting for $r$  from (\ref{eq:r-trans}) into (\ref{eq:cosmo-redshift}) gives
\begin{eqnarray}
\frac{r} {r'} &=& \frac{\left( r' + t v' \right) / r'} {\sqrt{1-t^2/ \tau^2}},  \label{eq:cosmo-redshift-1} \\
\nonumber \\
      &=& \frac{ 1 + t v' / r' } {\sqrt{1 - t^2 / \tau^2 } }.  \label{eq:cosmo-redshift-2}  
\end{eqnarray}
For a galaxy which is in the Hubble flow,
\begin{equation}
\frac{v'} { r'}  = \frac{1} {\tau}.  \label{eq:v'/r'=tau}
\end{equation}    
Substituting from (\ref{eq:v'/r'=tau})  into (\ref{eq:cosmo-redshift-2}), and along with (\ref{eq:cosmo-redshift}) 
yields
\begin{equation}
1 + z  =  \frac{ \lambda } { {\lambda}' }  =  \frac{r} {r'} 
         =  \frac{ 1 + t / \tau } {\sqrt{1 - t^2 / \tau^2 } }  = \sqrt{\frac{1 + t / \tau} {1 - t / \tau} }. 
             \label{eq:cosmo-redshift-final}
\end{equation}
Equation (\ref{eq:cosmo-redshift-final}) is the cosmological redshift of the wavelength of light measured 
between observers in frames $K$ and $K'$.  Inverting (\ref{eq:cosmo-redshift-final}) we get
\begin{equation}
\frac{t} {\tau}  = \frac{ \left( 1 + z \right)^2 - 1 } {  \left( 1 + z \right)^2 + 1 }.   \label{eq:t/tau_func(z)}
\end{equation}

\section{Dilation of cosmic time due to the expansion of space}

This transformation is similar to the lengthening of the wavelength of light from a distant galaxy 
by the factor $(1 + z)$. From the cosmological redshift relation (\ref{eq:cosmo-redshift-final})
and the fact that the periods $\tau_{\lambda}$ and $\tau'_{\lambda'}$ of a wave of light are related to the 
wavelengths $\lambda$ and $\lambda'$, respectively, by
\begin{eqnarray}
\tau_{\lambda}  = \frac{\lambda} {c},  \label{eq:tau_lambda} \\
\nonumber \\
\tau'_{\lambda}  = \frac{\lambda'} {c},  \label{eq:tau'_lambda'} 
\end{eqnarray}
we have from (\ref{eq:cosmo-redshift}),
\begin{equation}
\frac{ N \lambda / c } { {N \lambda'} / c } =   \frac{N \tau_{\lambda} } { N \tau'_{\lambda} }   
    = \frac{T} {T'} = 1 + z, 
   \label{eq:cosmo-redshift-dilation-1}
\end{equation}
where $T = \tau_{\lambda}$ and $T' = \tau'_{\lambda}$.  Then the cosmic time dilation from 
(\ref{eq:cosmo-redshift-dilation-1}) is given by
\begin{equation}
T = T' \left( 1 + z \right),  \label{eq:cosmo-redshift-dilation-2}
\end{equation}
where we assume that $T'$ is an arbitrary time interval in $K'$.

\section{Relativity of cosmic time}

Dividing  (\ref{eq:r-trans}) by (\ref{eq:v-trans}) we obtain the transformation
for the addition of cosmic time from $K'$ to $K$, analogous to the addition of velocities
in SR,
\begin{equation}
  t_2 = \frac{r} {v} = \frac{r' + t v'} {v' + t r'/\tau^2} = \frac{t'_1 + t} {1 + t  t'_1/\tau^2}, 
   \label{eq:t1-trans}
\end{equation}
where $t'_1=r'/v'$. The inverse transformation, from $K$ to $K'$ is obtained
by dividing (\ref{eq:r'-trans}) by (\ref{eq:v'-trans}) giving
\begin{equation}
  t'_1 = \frac{r'} {v'} = \frac{r - t v} {v - t r/\tau^2} = \frac{t_2 - t} {1 - t \, t_2/\tau^2}, 
   \label{eq:t2-trans}
\end{equation}
where $t_2=r/v$. We refer to (\ref{eq:t1-trans}) and (\ref{eq:t2-trans}) as the general cosmic time addition 
relations.  

Setting $t'_1 = \tau$ in (\ref{eq:t1-trans}) yields
\begin{equation}
t_2 = \frac{\tau + t} {1 + t  \tau/\tau^2} = \tau, 
   \label{eq:t1-trans-expan}
\end{equation}
implying that two cosmic times can never add up to more than $\tau$.  The identical result is obtained
from (\ref{eq:t2-trans})  for the time $t'_1$ when $t_2 = \tau$.

\subsection{Contraction of a small interval of cosmic time in the past
     \label{sec:contraction-of-time-interval} }

An increase in cosmic time $t$  by $\Delta{t'}$ in frame $K'$ at cosmic time $t$, where $\Delta{t'} \ll \tau$,
will have a value $t + \Delta{t}$ for the observer in $K$ at cosmic time $0$  given by the law 
of addition of cosmic times  (\ref{eq:t1-trans}) by setting $t'_1 = \Delta{t'}$ and $t_2 = t + \Delta{t}$,
\begin{equation}
  t +\Delta{t} = \frac{t + \Delta{t'}} {1 + t \Delta{t'} / \tau^2},  \label{eq:t + Delta-t_o} 
\end{equation}
which yields
\begin{eqnarray}
\Delta{t}  &=&    \frac{t + \Delta{t'}} {1 + t \Delta{t'} / \tau^2}   -  t \label{eq:Delta-t_o},  \\
\nonumber \\
                  &=&   \frac{ t + \Delta{t'} - t \left( 1 + t \Delta{t'} / \tau^2 \right) }  {1 + t \Delta{t'} / \tau^2},
                             \label{eq:Delta-t_o-expand}  \\
\nonumber \\ 
                     &\approx&  \Delta{t'} \left( 1 - t^2/\tau^2 \right),  \label{eq:Delta-t_o_approx}
\end{eqnarray}
since $\Delta{t'} \ll \tau$. This is a cosmic time contraction of a small time $\Delta{t'}$ in $K'$ at cosmic time $t$
measured by the observer in $K$\cite{hartnett-1}.   It implies that when viewed from the present epoch, 
a clock will appear to tick more slowly the further back it is in cosmic time.  
However, this does not alter the  physical constants like $G$, $\hbar$, or $e$, which remain constants.
This is analogous to the case in SR where a clock at the origin of a frame with relative velocity $v$ to 
the local frame, ticks at the rate $\delta{t'}$ in its own frame but appears to run more slowly at the rate 
$\delta{t'} / \sqrt{1 - v^2/c^2}$ when viewed from the local frame, but the physical constants do not vary. 

\subsection{Dilation of a small interval of cosmic time in the present}

There is a second kind of effect of time addition which is measured by the observer in $K'$ situated
at cosmic time $t$ from $K$.  Take the cosmic time lapse $\Delta{t} = (t+\Delta{t}) - t$ recorded 
in $K$ where $\Delta{t} \ll \tau$. What is the observation of that time lapse for
the observer in $K'$?  In other words, in $K'$ what is the difference of the later time $t+\Delta{t}$ 
with respect to the earlier time $t$? The time transformation we use for the $K'$ frame is  from 
(\ref{eq:t2-trans}) by setting $t'_1 = \Delta{t'}$, $t_2 = t + \Delta{t}$ giving
\begin{eqnarray}
\Delta{t'}  &=&  \frac{\left( t + \Delta{t} \right) - t} {1 - t \,  \left( t + \Delta{t} \right) /\tau^2},
      \label{eq:Delta-t-restfram-1}  \\
\nonumber \\
     &\approx&  \frac{  \Delta{t} } { 1 - t^2/\tau^2 },  \label{eq:Delta-t-restfram-2} 
\end{eqnarray}
since $\Delta{t} \ll \tau$. This is a cosmic time dilation of a small time interval in frame $K$ measured in $K'$.
It infers that a short time interval at the present epoch corresponds to a larger time interval further
back in cosmic time.  Note that if $t + \Delta{t} = \tau$ then (\ref{eq:Delta-t-restfram-1}) gives 
$\Delta{t'} = \tau$ so that we never get a time greater than $\tau$ as long as we add times $\le \tau$.

\section{Total cosmic time transformation due to the expansion of space and the additon of cosmic times}

\begin{figure}[htb!]
\fbox{\includegraphics[viewport=0 135 275 420,keepaspectratio,clip=true]{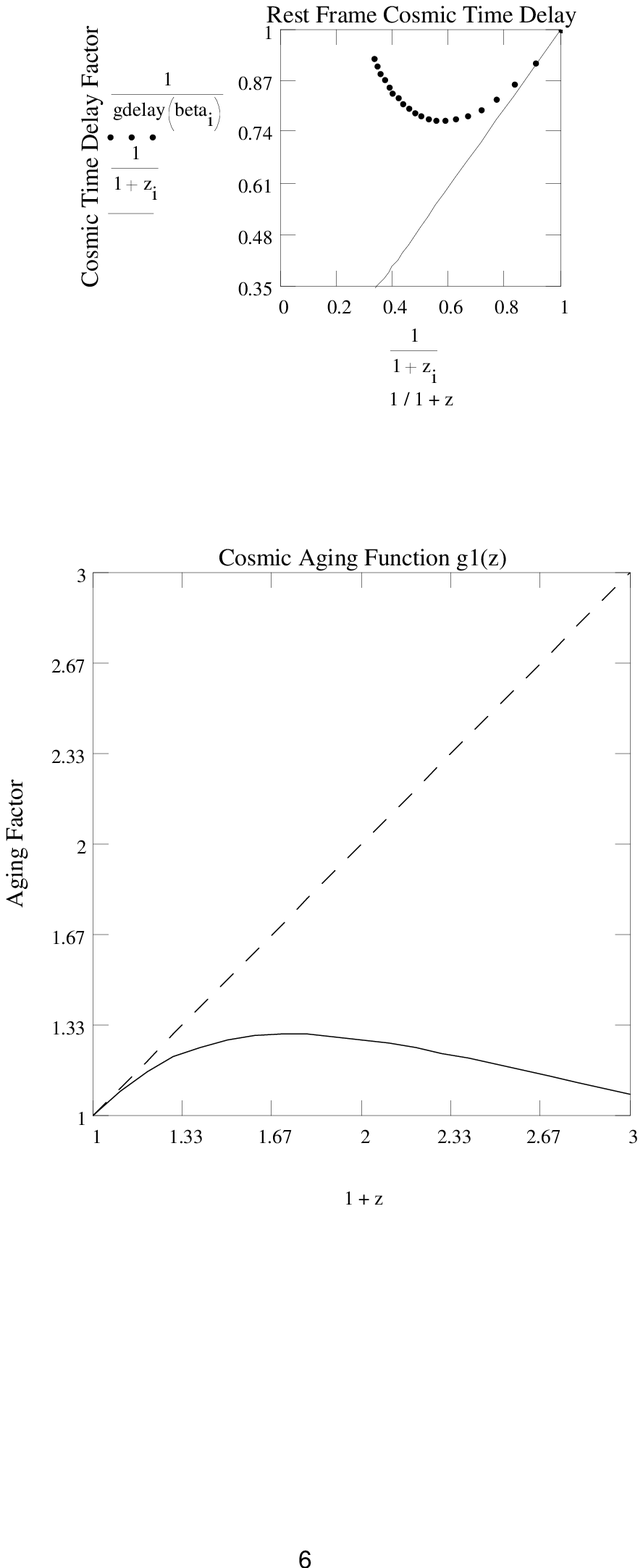}}  
\setlength{\fboxrule}{1.5pt}
\vspace*{8pt}
\caption{Cosmic aging function $g_1(z) = 4 (1+z)^3/[(1+z)^2 + 1]^2$,  $0 < z < 2$, solid line.
The dashed line is $1 + z$. }
\label{fig:g_1(z)-0-to-2}
\end{figure}

Combine  the two transformations by taking the product of time dilation (\ref{eq:cosmo-redshift-dilation-2}) 
due to the expansion of the universe with time contraction (\ref{eq:Delta-t_o_approx}) due to the addition of 
cosmic times. The total elapsed cosmic time $\Delta{t}$ observed by the observer in $K$ at cosmic time $0$
for a small time change $\Delta{t'} \ll \tau$ in frame $K'$ at cosmic time $t$ is given by 
\begin{equation}
\Delta{t} =  \left( 1 + z \right)  \left( 1 - t^2/\tau^2 \right) \Delta{t'}  =  g_1( t ) \Delta{t'}, 
        \label{eq:cosmo-transfrm-1-kind}
\end{equation}
where by substituting for $1 + z$ from (\ref{eq:cosmo-redshift-final}), the cosmic aging function $g_1(t)$ 
is defined by
\begin{equation}
 g_1\left( t \right)  =   \left(\sqrt{\frac{1 + t / \tau} {1 - t / \tau} } \right) \left( 1 - t^2/\tau^2 \right).  \label{eq:f(t/tau)}
\end{equation}
Substituting for $t / \tau$ from (\ref{eq:t/tau_func(z)}) in terms of redshift $z$ into (\ref{eq:f(t/tau)})
yields,
\begin{equation}
g_1\left( z \right) = \frac{ 4 \left( 1 + z \right)^3 } { \left[ \left( 1 + z \right)^2 + 1 \right]^2 }.    \label{eq:g_1(z)-define}
\end{equation}
The cosmic aging function $g_1(t) > 1$ for $t / \tau < 0.8392$, which gives a time dilation,
and $g_1(t) < 1$ for $t / \tau > 0.8395$, which corresponds to a time contraction. 
The maximum occurs at $t / \tau = 1/2$ where $g_1(\tau/2) = 1.299$ which, from (\ref{eq:cosmo-redshift-final}), 
corresponds to a redshift $z \approx 0.732$. Fig. \ref{fig:g_1(z)-0-to-2} is a  plot of the cosmic aging function $g_1(z)$.

\section{Distances}

We use the distance relation (\ref{eq:r-with-any-Omega}) with the density $\Omega = \Omega_M$, giving
\begin{equation}
r  =  \frac{ c \tau \, {\rm sinh}\left(  \left( v / c \right)  \sqrt{ 1 - \Omega_M } \right) }  
                {\sqrt{ 1 - \Omega_M } }
     \label{eq:r-distance-Omega(t)}
\end{equation}
where $\Omega_M$ is the average mass density parameter at the present epoch.

In CGR the luminosity distance relation $D_L$ includes the contraction of a small interval of cosmic time
in the past, (\ref{eq:Delta-t_o_approx}).  It enters through the concept that the energy measured over an
interval of time  $\Delta{t'}$ from a source with luminosity $L$ at rest in frame $K'$ at cosmic time $t$,
radiates a proportional quantity of energy $E$ measured by the observer in $K$ given by
\begin{equation}
E \propto L \left( 1 - t^2 / \tau^2 \right) \left( 1 + z \right) \Delta{t'}  =  L  \, g_1\left(t \right) \Delta{t'},
   \label{eq:Energy-luminosity}
\end{equation}
which from (\ref{eq:f(t/tau)})  is the cosmic aging function $g_1(t)$ operating on the time interval 
$\Delta{t'}$.
Using (\ref{eq:Energy-luminosity})  as the form for the energy of the source in the derivation\cite{hartnett-1}, 
the luminosity distance $D_L$ is given by
\begin{equation}
D_L  =   \frac{ r \left( 1 + z \right) } { \sqrt{ 1 - t^2 / \tau^2 } },  \label{eq:Lum-dist}
\end{equation}
which is a factor $( 1 - t^2 / \tau^2 )^{-1/2}$ of the standard form $D_L = r (1 + z)$, implying that
in CGR sources appear less luminous and thus further away due to the relativity of cosmic time.
Substituting for distance $r$ from (\ref{eq:r-distance-Omega(t)}) and for $1 + z$ from (\ref{eq:cosmo-redshift-final}),
and simplifying, (\ref{eq:Lum-dist}) becomes
\begin{equation}
D_L\left( t \right) =    \frac{ c \tau \, {\rm sinh}\left( \left( v / c \right) \sqrt{ 1 - \Omega_M} \right) }
      { \left( 1 - t / \tau \right) \sqrt{ 1  - \Omega_M } }. \label{eq:Lum-dist-tbytau} 
\end{equation}
In practice we make the substitution $t/\tau = v/c$, (Ref. \cite[B.4.2]{carmeli-0}), for which another derivation is given in 
appendix \ref{ap:dv/dt}.

\section{Distance data fitting}

We apply the luminosity distance relation (\ref{eq:Lum-dist-tbytau}) plus a calculated best fit fixed offset $a_{off}$
to get the apparent magnitude $m(z)$ of a distant luminous source,
\begin{equation}
m\left( z \right) = 5 \log\left[D_L \left( z \right) \right] + M_B + a_{off},
            \label{eq:Lum-dist-z-aoff}
\end{equation}
where we apply (\ref{eq:t/tau_func(z)}) to convert from $t/\tau$ to $z$ in obtaining $D_L ( z )$.
We use the CGR standard value\cite{oliveira-0} of $c \tau = 4158  \, {\rm Mpc}$. In practice the source absolute 
magnitude $M_B$ is absorbed into the value of the offset $a_{off}$.  We include in our analysis data from both
SNE-Ia and GRB studies. The SNE-Ia data come from the Supernova Cosmology Project SCP Union2.1 
dataset\cite{SCPUnion-1} of $580$ SNe-Ia magnitudes and errors up to $z<1.5$.  The GRB distance data of $69$ 
burst events come from\cite{schaefer-1} which are selected events up to $z < 7$ from website data provided 
by\cite{greiner-1}.   For the $m_j$ observed magnitudes and $merr_j$ respective errors, the $\chi^2$ for the fit 
is defined by
\begin{equation}
\chi^2  =  \sum^N_{j=1}{ \left[ \frac{m_j - m\left(z_j \right) } { merr_j } \right]^2}.
    \label{eq:chi-sqr-SNe}
\end{equation}
The reduced chi-squared $\chi_{red}^2$ is given by
\begin{equation}
\chi^2_{red}  =  \frac{\chi^2} {\left( N - k -1 \right)},
\end{equation}
where $N$ is the number of data samples and $k$ is the number of fitting parameters.

For the ${\Lambda}CDM$ model the luminosity distance relation is given by
\begin{equation}
D_{L{\lambda}cdm} \left( z \right) = c \tau \left( 1 + z \right) \int^z_0{\frac{du} {\sqrt{\Omega_M \left( 1 + u \right)^3 + \Omega_{\Lambda} } }},
   \label{eq:DLlcdm-def}
\end{equation}
where $\Omega_M + \Omega_{\Lambda} = 1$ for flat space\cite{weinberg-1}. 

\begin{figure}[htb!]
\fbox{\includegraphics[viewport=-2 275 400 635,keepaspectratio,clip=true]{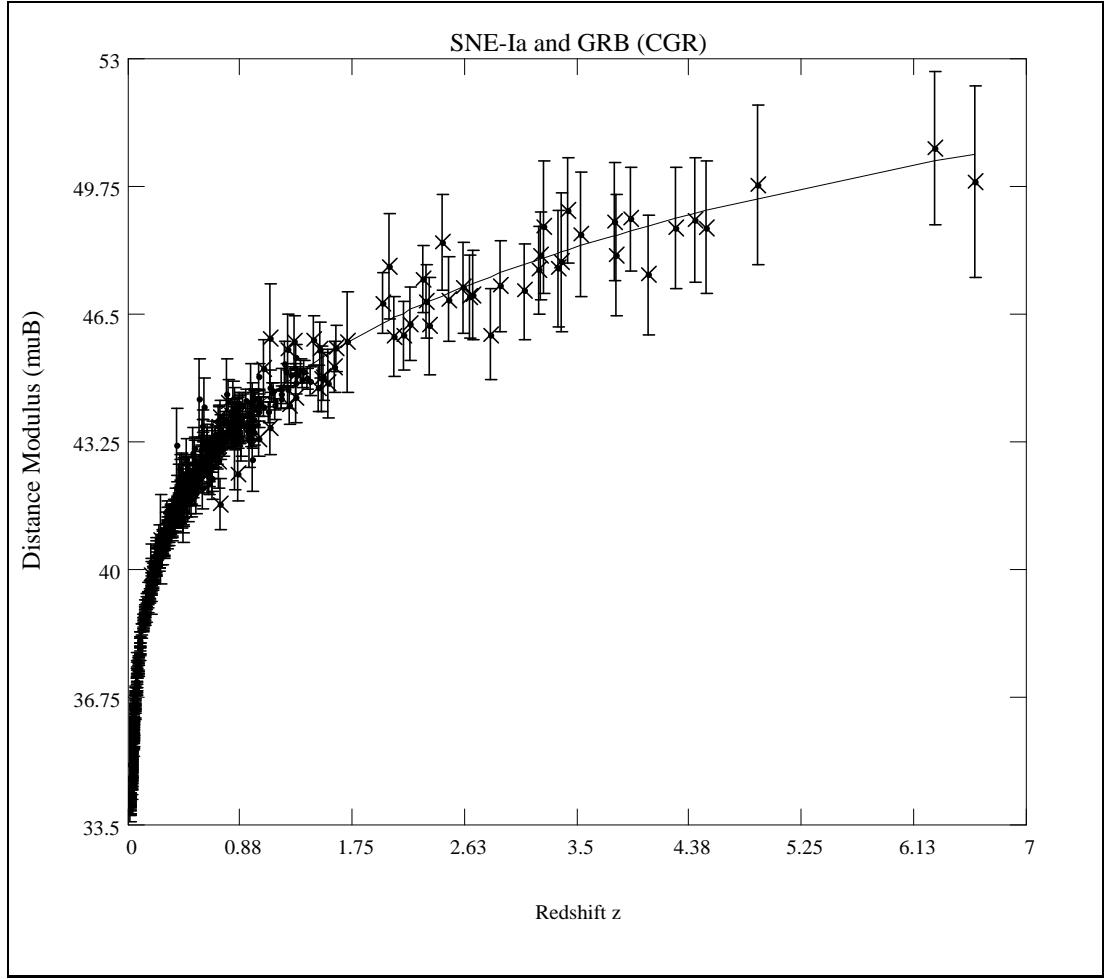}} 
\vspace*{8pt}
\caption{SCP Union 2.1 SNe-Ia data\cite{SCPUnion-1} (filled circles)  and GRB\cite{schaefer-1} (dotted x's). 
The solid line is for CGR $DL(z)$ from (\ref{eq:Lum-dist-z-aoff}).
The CGR standard value was used for $1 / \tau = h  =  72.2 \pm 0.84 \, {\rm km / s / Mpc}$.
The parameters for the CGR model were the calculated best fit values, with  mass density 
$\Omega_M = 0.800$,  offset  $a_{off} = 0.140$ and GRB conversion factor $\gamma_1=1.044442$.
Magnitude and magnitude errors both were converted to the CGR model (\ref{eq:grb-lcdm-to-cgr}).
For the fit to 649 data points with 3 parameters the reduced $\chi^2 = 1.001182$.  
\label{fig:DL(z)-SCPUnion+GRB(CGR)} }
\end{figure}

\begin{figure}[htb!]
\fbox{\includegraphics[viewport=-3 290 230 580,keepaspectratio,clip=true]{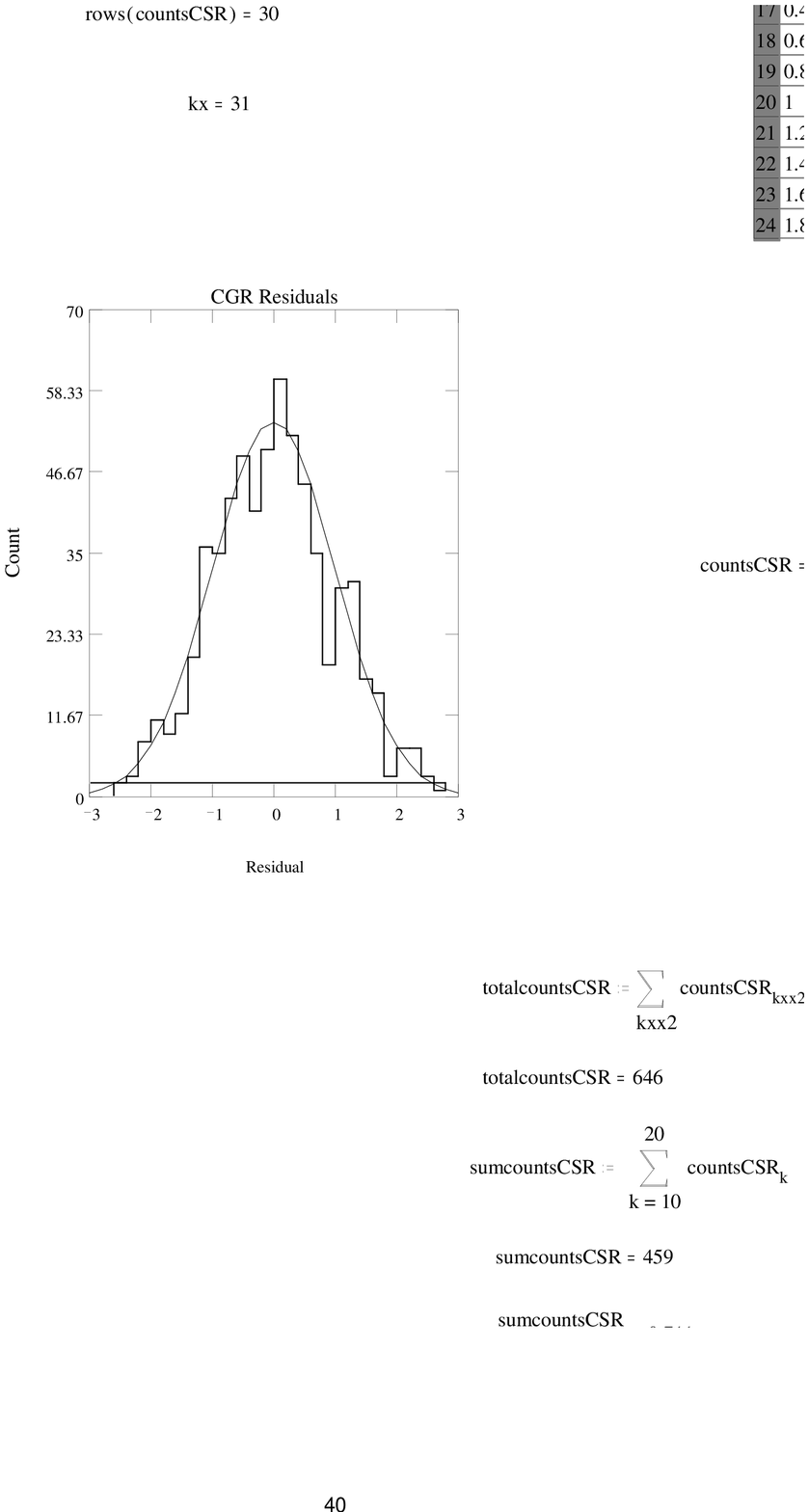}} 
\vspace*{8pt}
\caption{Histogram of CGR residuals $m_k - m( z_k )$ with $DL(z_k)$ from 
(\ref{eq:Lum-dist-z-aoff}) with redshift $z_k$ from the SCP Union 2.1 SNe-Ia data\cite{SCPUnion-1} and GRB 
data\cite{schaefer-1}, for calculated best fits with mass density $\Omega_M = 0.800$, offset $a_{off} = 0.140$
and $\gamma_1 = 1.044442$. 
The solid line is a standard Gaussian with mean $\mu = 0$ and standard deviation $\sigma = 1$ and the amplitude
is scaled by the factor $A = 135$ to give a good ``by eye" fit to the histogram.  }
\label{fig:CGR-Histogram}
\end{figure}

Although SNE-Ia data are independent of any particular cosmology, this is not so for GRB data which must be 
calibrated with a specified cosmological model.  This is because SNE-Ia have nearby sources to use for calibration,
but for GRB there are no nearby sources for this purpose. The original GRB data set was calibrated 
with the ${\Lambda}CDM$ cosmology. To recalibrate the GRB data for the CGR cosmological model\cite{hartnett-0}
we use the relation
\begin{equation}
M_{cgr} \left( z \right) = M_{{\lambda}cdm} \left( z \right)
          + {\rm log} \left[\frac{  \gamma_1  DL\left( z \right) }
                                                 { DL_{{\lambda}cdm} \left( z \right) } 
                 \right],    \label{eq:grb-lcdm-to-cgr}
\end{equation}
where $M_{{\lambda}cdm}$ is the original GRB distance modulus calibrated with the ${\Lambda}CDM$ model,
$M_{cgr} ( z )$ is the distance modulus calibrated for the CGR model and $\gamma_1$ is a conversion factor.   The 
magnitude errors were also converted using (\ref{eq:grb-lcdm-to-cgr}).
We determined a good fitting value of $\gamma_1 = 1.044442$ which was used to convert the GRB distance moduli 
for all fits to the CGR model.  The converted GRB data were combined with the SNE-Ia data to form the 
complete data set. For the CGR model, the number of parameters $k = 3$ for $\Omega_M$, $a_{off}$ and 
$\gamma_1$. The parameter $\tau$ is fixed in this analysis, so the number of degrees 
of freedom $N_{dof} = N - k - 1 = 645$, the same as for the ${\Lambda}CDM$ model. 
The best fit for the CGR model is shown in Table \ref{tb:ReducedChiSq} for $\Omega_M = 0.800 \pm 0.080$
 (a conservative estimate of $\pm 10 \%$ error) with offset $a_{off} = 0.140$ and $\gamma_1 = 1.044442$ 
having a reduced chi-squared $\chi^2/ N_{dof} = 1.001182$ for $N_{dof} = 645$.
The results are shown in Fig. \ref{fig:DL(z)-SCPUnion+GRB(CGR)}. 
Figure \ref{fig:CGR-Histogram} shows the histogram of the normalized residual errors 
for the fit.  The solid curve is a Gaussian with mean $\mu = 0$ and standard deviation $\sigma = 1$ with 
an amplitude $A=135$ estimated ``by eye" to give a close fit to the histogram.  The fit appears 
good\cite{schaefer-1,hartnett-0,oztas-1}.

\begin{figure}[htb!]
\fbox{\includegraphics[viewport=0 105 395 460,keepaspectratio,clip=true]{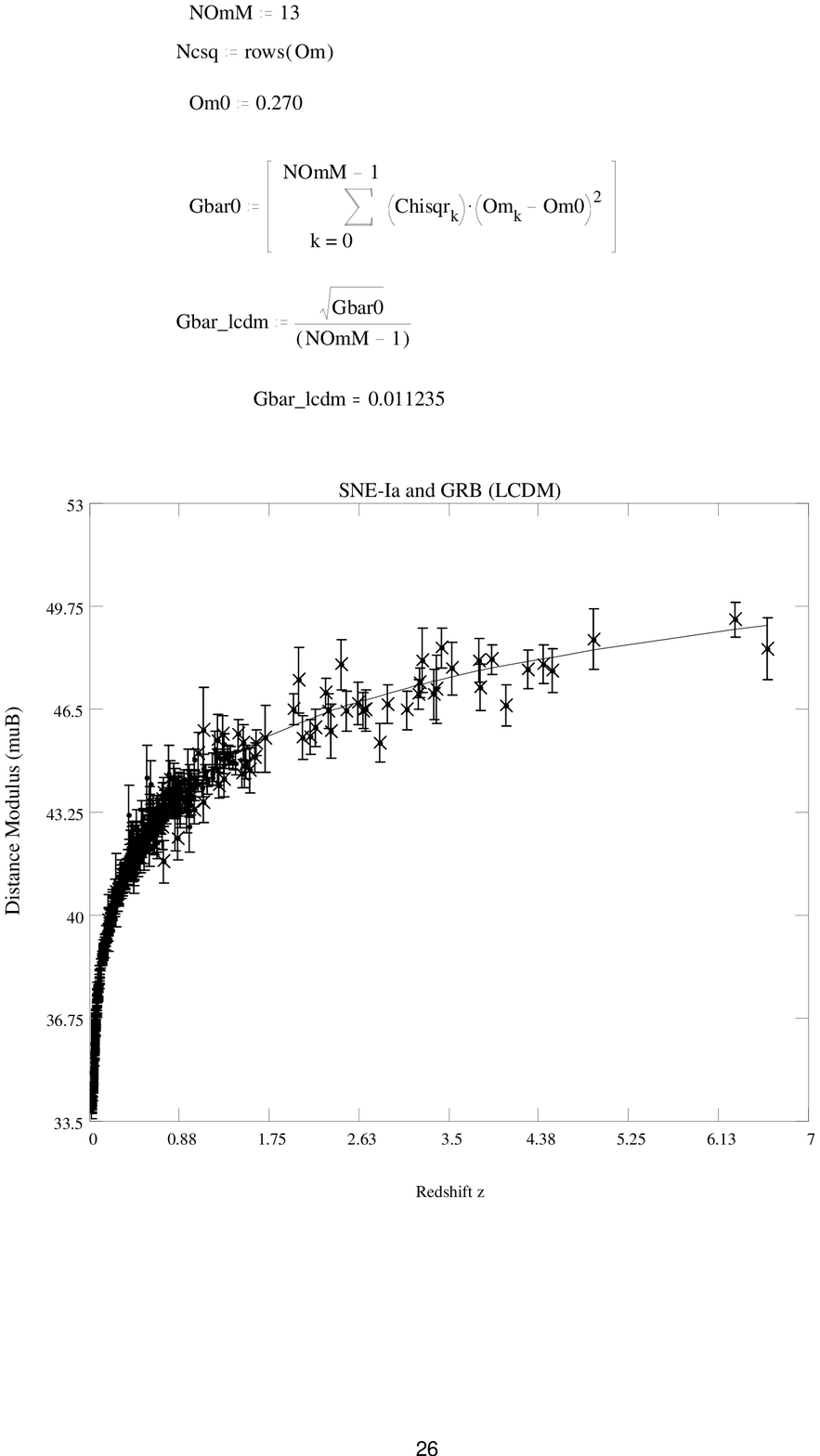}} 
\vspace*{8pt}
\caption{SCP Union 2.1 SNe-Ia data\cite{SCPUnion-1} (filled circles)  and GRB\cite{schaefer-1} (dotted x's).
The solid line is for $DL_{{\lambda}cdm}(z)$ from (\ref{eq:DLlcdm-def}). 
The CGR standard value was used for $1 / \tau = h  =  72.2 \pm 0.84 \, {\rm km / s / Mpc}$.
For the ${\Lambda}CDM$ model, with fixed $\Omega_M = 0.270$ and fixed $\Omega_{\Lambda} = 0.730$,
the calculated best fit value for offset $a_{off} = 0.060$.
For the fit to 649 data points with 3 parameters the reduced $\chi^2 = 0.985111$. }
\label{fig:DL(z)-SCPUnion+GRB(LCDM)}
\end{figure}

\begin{figure}[htb!]
\fbox{\includegraphics[viewport=30 285 265 580,keepaspectratio,clip=true]{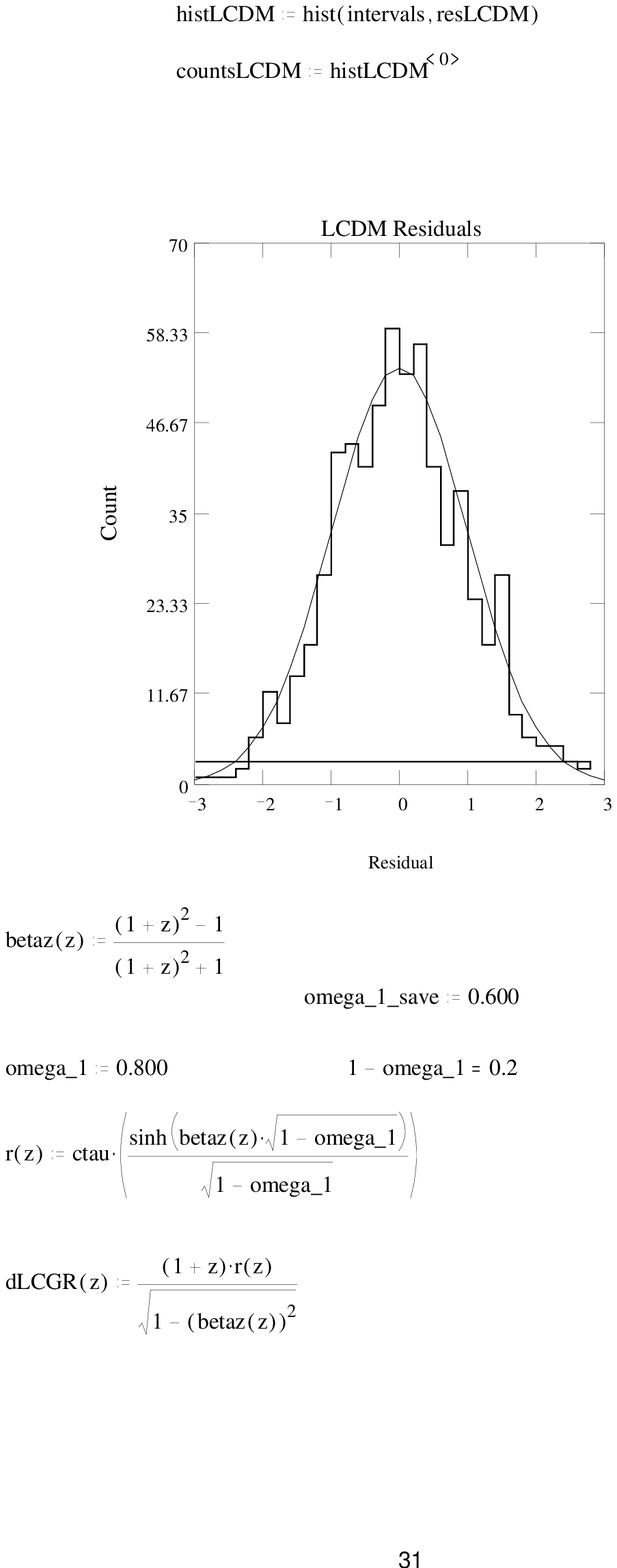}} 
\vspace*{8pt}
\caption{Histogram of ${\Lambda}CDM$ residuals $m_k - m( z_k )$ for $DL_{{\lambda}cdm}(z_k)$ from 
(\ref{eq:DLlcdm-def}) with redshift $z_k$ from the SCP Union 2.1 SNe-Ia data\cite{SCPUnion-1} and GRB 
data\cite{schaefer-1}, for fixed mass density $\Omega_M = 0.270$ and fixed dark energy mass density
$\Omega_{\Lambda} = 0.730$, with a calculated best fit offset $a_{off} = 0.060$.
The solid line is a standard Gaussian with mean $\mu = 0$ and standard deviation $\sigma = 1$ and the amplitude
factor $A = 135$ comes from the CGR histogram.  }
\label{fig:LCDM-Histogram}
\end{figure}

We fit the ${\Lambda}CDM$ cosmology to the original GRB data set, which was calibrated  with 
$\Omega_M = 0.270$ and  $\Omega_{\Lambda} = 0.730$.  We assume $k=3$ parameters in the model, 
for $\Omega_M$,  $\Omega_{\Lambda}$  and  $a_{off}$.  We use a fixed Hubble constant of  
$H_0 = 1 / \tau = 72.2 \, {\rm km \, s^{-1} \, Mpc^{-1}}$. The best fit occurred for offset 
$a_{off} = 0.060$.  Fig. \ref{fig:DL(z)-SCPUnion+GRB(LCDM)} shows the Hubble Diagram for 
the fit of $D_{L{\lambda}cdm} ( z )$ to the combined data, with the GRB moduli used unaltered from the data set.
The parameter $H_0$ is fixed in this analysis, so the number of degrees of freedom is the same as for the
CGR model. The reduced chi-squared $\chi^2 / N_{dof} = 0.985111$. 
Figure \ref{fig:LCDM-Histogram} shows the histogram of 
the normalized residual errors for the fit.  
The solid curve is a Gaussian with mean $\mu = 0$ and standard deviation $\sigma = 1$ with an amplitude $A=135$ 
taken from the CGR histogram. 

Under the reduced chi-squared statistical model,  the ideal reduced $\chi^2=1$ with the errors
distributed normally (Gaussian) about $\mu = 0$ with $\sigma = 1$.  The model with a reduced $\chi^2$ 
which is closest to $1$ is preferred. Models with reduced $\chi^2 > 1$ are deemed to have too few parameters,
so "under fit" the data.  Models with  reduced $\chi^2 < 1$ are deemed to have too many  
parameters and thus "over fit" the data.

We see that the CGR model under fits the data by $\approx 0.001182$, while the ${\Lambda}CDM$ model
over fits the data by  $\approx 0.014889$.  For this analysis, the CGR model is $\approx 10$ times better at
fitting the combined data.  However, considering that we did not vary the mass densities $\Omega_M$ and 
$\Omega_{\Lambda}$, nor the Hubble parameter $H_0$, when fitting with the ${\Lambda}CDM$ model,
we make this conclusion as mainly a statement of our confidence in the CGR model. A more rigorous analysis
of the fitting operation would be required.

\subsection{Dark matter and the X particle hypothesis}

Assume that the mass density $\Omega_M = \Omega_B + \Omega_D$,
that is, composed of baryonic matter $\Omega_B$ and cold dark matter $\Omega_D$.  For 
$\Omega_B h^2 \approx 0.020 \pm 0.002$ ($95 \, \%$ confidence level)\cite{burles-1},
with the CGR value $h=0.722$, this gives
\begin{equation}
 \Omega_B \approx 0.038 \pm 0.004.  \label{eq:OmB-value}
\end{equation}
Then, for $\Omega_{Mcgr} = 0.800 \pm 0.080$ and $\Omega_{M{\lambda}cdm} = 0.270 \pm 0.013$,
 the CGR model gives a dark matter density of 
\begin{equation}
\Omega_{Dcgr} \approx 0.762 \pm  0.084,  \label{eq:Omega-dark-crg}
\end{equation}
compared to the ${\Lambda}CDM$ model  which gives a dark matter density of 
\begin{equation}
\Omega_{D{\lambda}cdm} \approx 0.232 \pm 0.017.  \label{eq:Omega-dark-lcdm}
\end{equation}
In an extension to the standard model (SM), a hypothetical $X$ particle\cite{davoudiasl-1} is theorized to exist,
having 2 species, $X_1$ and $X_2$ (and their conjugate species $\bar{X}_1$ and $\bar{X}_2$.)
These particles were generated non-thermally during the early universe.
The $X_1$ decays either
into a visible three quark state ($UDD$), or the hidden state ($Y, \Phi$), with each state having baryon 
number $+1$.  The conjugate state $\bar{X}_1$ decays to the visible three quark state ($\bar{U}\bar{D}\bar{D}$) 
or the hidden state ($\bar{Y}, \Phi^{*}$), with each state having baryon number $-1$.
All of the dark matter today, in this extended model, is theorized to be composed entirely of hidden particle states
($\bar{Y}, \Phi^{*}$). Rare processes can transfer baryonic number from the hidden sector to the 
visible sector through inelastic scattering of anti-baryonic dark matter states ($\bar{Y}, \Phi^{*}$),  annihilating baryons 
in the visible sector. The cosmic abundance of remnant $\bar{Y}$ and $\Phi^{*}$  particles,  with densities given by 
$n_{\bar{Y}}$ and  $n_{\Phi^{*}}$, respectively,  is the same as the baryon density $n_B$ in the universe today, 
and thus would have the same abundance ratio as $\eta_B$, the baryon to photon ($n_{\gamma}$) ratio. That is,
\begin{equation}
\frac{n_{\bar{Y}}} {n_{\gamma} } = \frac{n_{\Phi^{*}}} {n_{\gamma} } = \frac{n_B} {n_{\gamma} } 
      = \eta_B  \approx 6 \, \times \, 10^{-10},       \label{eq:n_X/n_gamma}
\end{equation}
for a baryon density $\Omega_B \approx 0.04$.
Further details of this extension to SM is beyond the scope of this paper.
From Ref. (\cite{davoudiasl-1}, (10))  we can relate the ratio of the density of dark matter (anti-baryonic) to 
baryonic matter in the universe to the ratio of the rest masses of the $\bar{Y}$,  $\Phi^{*}$ and proton by
\begin{equation}
\frac{\Omega_D} {\Omega_B}  = \frac{ m_{\bar{Y}}  + m_{\Phi^{*}} } { m_p } \approx  \frac{2 w \, m_p} {m_p}  = 2 w,
        \label{eq:OmDark/OmB}
\end{equation}
where $m_{\bar{Y}}$ and  $m_{\Phi^{*}}$ are, respectively, the $\bar{Y}$ and $\Phi^{*}$ particle rest masses 
and $m_p$ is the proton rest mass and we assumed that $m_{\bar{Y}}  \approx m_{\Phi^{*}}  \approx w \, m_p$.
For the CGR model, (\ref{eq:OmDark/OmB}) yields
\begin{equation}
\frac{\Omega_{Dcgr}} {\Omega_B}  = \frac{0.762 \pm 0.084} {0.038 \pm 0.004} = 2 w_{cgr},  
      \label{eq:CGR-w-calc}
\end{equation}
which implies
\begin{equation}
w_{cgr} = 10.43 \pm 2.18.
\end{equation}
This gives rest mass energies for the $\bar{Y}$ and $\Phi^{*}$ particles of
\begin{equation}
m_{{\bar{Y}}cgr} c^2 \approx m_{{\Phi^{*}}cgr} c^2  \approx  \left( 9.79  \pm  0.47 \right) {\rm GeV}.  
      \label{eq:mY_mPhi-mass-CGR}
\end{equation}
For the  ${\Lambda}CDM$ model,  (\ref{eq:OmDark/OmB}) yields
\begin{equation}
\frac{\Omega_{D{\lambda}cdm}} {\Omega_B}  = \frac{0.232 \pm 0.017} {0.038 \pm 0.004} =  2 w_{{\lambda}cdm}, 
      \label{eq:LCDM-w-calc}
\end{equation}
which implies
\begin{equation}
w_{{\lambda}cdm} = 2.61 \pm 0.55.
\end{equation}
This gives rest mass energies for the $\bar{Y}$ and $\Phi^{*}$ particles\cite{davoudiasl-1} of
\begin{equation}
m_{{\bar{Y}}{\lambda}cdm} c^2 \approx m_{{\Phi^{*}}{\lambda}cdm} c^2  \approx  \left( 2.45  \pm  0.47 \right) {\rm GeV}.  
        \label{eq:mY_mPhi-mass-LCDM}
\end{equation}
In both cases above we have tried to account for the constraint that $\mid m_{\bar{Y}} - m_{\Phi^{*}} \mid < m_p + m_e$, where
$m_e$ is the electron mass, by restricting the range of values to be within $\pm 0.47 \, {\rm GeV}$. This may be only
an approximate treatment, at best.

\begin{table}
\caption{CGR vs. ${\Lambda}CDM$ model performances with reduced $\chi^2 / (n-k-1) $. The number of samples $n = 649$,
and number of parameters $k=3$ for both models which gives $n-k-1 = 645$.  The CGR row showing the best fit is marked 
with a (*). Refer to the text for an explanation of the best fitting model. The $\Lambda$CDM model, with fixed $\Omega_M$
and fixed $\Omega_{\Lambda}$ has a single best fit for $a_{off}$ based on the original data.}
\begin{tabular}{| r | c | c | c | c|}
\hline
Model                   & $\Omega_M$  &  $a_{off}$     &  $\chi^2$     &  $\chi^2 / 645$   \\
\hline
CGR                     &  $0.650$         &    0.140          &  640.656     &   $0.993265$   \\
\hline
CGR                     &  $0.700$         &    0.140          &  642.143      &   $0.995571$   \\
\hline 
CGR                     &  $0.750$         &    0.140          &  643.844       &   $0.998208$   \\
\hline
*CGR                     &  $0.800$         &    0.140          &  645.762       &   $1.001182$   \\
\hline
CGR                     &  $0.850$         &    0.143          &  648.263       &   $1.005059$   \\
\hline
CGR                     &  $0.900$         &    0.145          &  650.815       &   $1.009015$   \\
\hline
$\Lambda$CDM   &  $0.270$         &    0.060          &  635.397       &   $0.985111$   \\
\hline
\end{tabular}
\label{tb:ReducedChiSq}
\end{table}

\section{Time dilation in SNe-Ia light curves}

We now consider two SNe-Ia (SNe) light curve experiments\cite{goldhaber-1,blondin-1}.
Common to both experiments 
is the stretch of the SNe light curve interval for each distant source when compared to the standard nearby (local) 
source.  Because these studies rely on a model for what the light curve looks like in the rest frame of the source
SNe,  we will require the use of both kinds of cosmic time additions described above.
The light curve time interval $\Delta{t'_{obsspec}}$  from the distant SNe will be observed to have a time transformation
which is a combination of cosmic time addition in the past combined with the time effect of the expansion of space,
having the observed value $\Delta{t_{obs}}$ which is given by (\ref{eq:cosmo-transfrm-1-kind}),
\begin{equation}
\Delta{t_{obs}}  =  \Delta{t'_{obsspec}} \left( 1 + z \right) \left( 1 - t^2 / \tau^2 \right).  \label{eq:delta-t_obs}
\end{equation}
In CSR this is the time interval that is observed from a source light curve or any time varying phenomenon
at a cosmic time $t$ in the past.  

On the other hand, we can describe the light curve recorded by the observer in the
frame $K'$ at cosmic time $t$ relative to our local observer in $K$ at cosmic time $0$.  A light curve time duration
of $\Delta{t_{spec}}$ in $K$, from (\ref{eq:Delta-t-restfram-2}) for a cosmic time addition in the present,
corresponds to the value $\Delta{t'_{spec}}$ in $K'$ given by
\begin{equation}
\Delta{t'_{spec}} = \frac{ \Delta{t_{spec}} } { \left( 1 - t^2 / \tau^2 \right) }.  \label{eq:delta-t_spec}
\end{equation}
If we make the assumption that
\begin{equation}
\Delta{t'_{obsspec}}  =  \Delta{t'_{spec}},     \label{eq:delta-t-obsspec=delta-t-spec}
\end{equation}
then combining (\ref{eq:delta-t_obs})  with (\ref{eq:delta-t-obsspec=delta-t-spec}) yields
\begin{equation}
\Delta{t_{obs}}  = \Delta{t_{spec}}  \left( 1 + z \right).  \label{eq:delta-t-obs-t-spec-final}
\end{equation}
It is evident that CSR, assuming (\ref{eq:delta-t-obsspec=delta-t-spec}),
is consistent with the time dilation reports showing effects of cosmic aging equivalent with $1 + z$ for redshift $z$.  
However,  to offer a different perspective on cosmic time transformation, we will show plots of the ratio
$\Delta{t_{obs}} / \Delta{t'_{obsspec}}$
given by (\ref{eq:delta-t_obs}), instead of $\Delta{t_{obs}} / \Delta{t_{spec}}$ which was used in those reports.

\begin{figure}[htb!]
\fbox{\includegraphics[viewport=-5 145 370 420,keepaspectratio,clip=true]{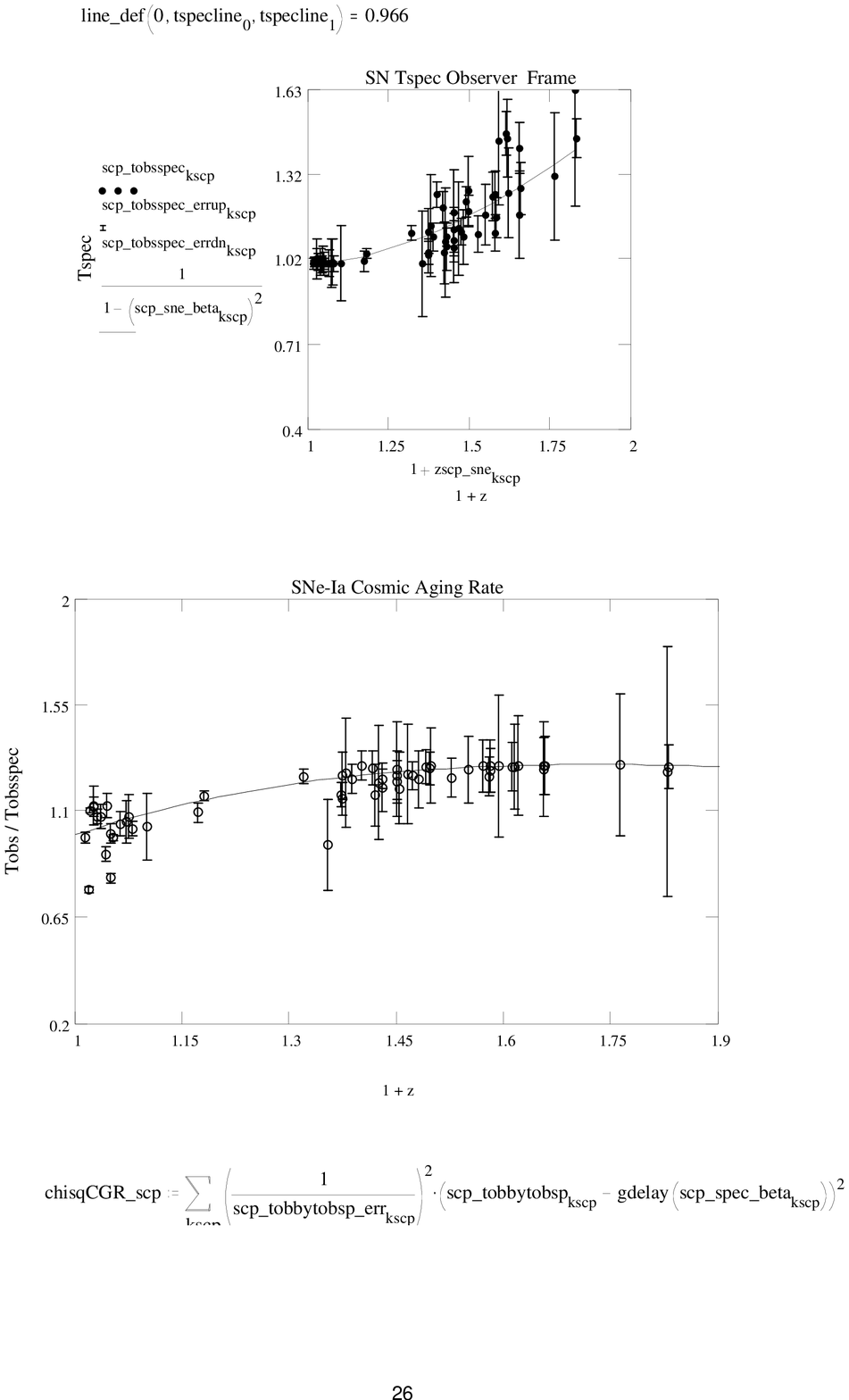}} 
\vspace*{8pt}
\caption{Cal\'{a}n/Tololo low $z$ and SCP high $z$ SNe-Ia light curve time intervals\cite[Goldhaber, et al.]{goldhaber-1}. 
Open circles are $\Delta{t_{obs}} / \Delta{t'_{obsspec}}$ for each of the 58 SNe-Ia.  
Error bars are computed from the $\sigma_w$ data errors. 
The solid line is the cosmic aging function $g_1(z)$ of  (\ref{eq:g_1(z)-define}). The reduced $\chi^2$ for the fit of $g_1(z)$
to the data is $\chi^2 / 57 = 10.8$. }
\label{fig:SCP-Goldhaber-2}
\end{figure}

For the SNe data from\cite{goldhaber-1} the light curve agings are given as the light curve width
$w$ and the error in the width $\sigma_w$,  obtained directly from \cite[Table 1]{goldhaber-1} for the SCP high $z$ SNe 
and from \cite[Table 3]{goldhaber-1} for the Cal\'{a}n/Tololo low $z$ SNe. 
Since the goal of the experiment was to normalize each light curve to a single standard light curve, we will
assume that the equivalent local rest frame time is $\Delta{t_{spec}} = 1$ which implies from (\ref{eq:delta-t_spec})
that $\Delta{t'_{spec}} = 1 / ( 1 - t^2 / \tau^2 )$.  The reduced observed quantity is $w$ and we will assume
using (\ref{eq:delta-t-obs-t-spec-final}),
\begin{equation}
\Delta{t_{obs}} = \Delta{t_{spec}} \left( 1 + z \right) = \left( 1 + z \right)   =  w.  \label{eq:delta-t-obs=w}
\end{equation}
We will use (\ref{eq:delta-t-obs=w}) to acquire the redshift $z$ and hence $t / \tau$ from the light curve
rather than using the redshift from the host galaxy.  The quantity we use is the ratio 
$\Delta{t_{obs}} / \Delta{t'_{obsspec}}$ from (\ref{eq:delta-t_obs}),
\begin{equation}
\frac{ \Delta{t_{obs}} }  { \Delta{t'_{obsspec}} }  =  \left( 1 + z \right) \left( 1 - t^2 / \tau^2 \right).
      \label{eq:Tobs_by_Tobsspec}
\end{equation}
The plotted data are shown in Fig. \ref{fig:SCP-Goldhaber-2}.  For data points with $w < 1$ the redshift was set 
to $z=0$. The plot shows the reduction of apparent light curve
aging at higher redshift. This is the effect which would be seen in the observed light curve without scaling by the
rest frame aging rate. The reduced chi-squared  is  $\chi^2 / 57 = 10.8$ for the data fitted to the cosmic aging 
rate $g_1(z)$, (\ref{eq:g_1(z)-define}).

\begin{figure}[htb!]
\fbox{\includegraphics[viewport=0 170 380 450,keepaspectratio,clip=true]{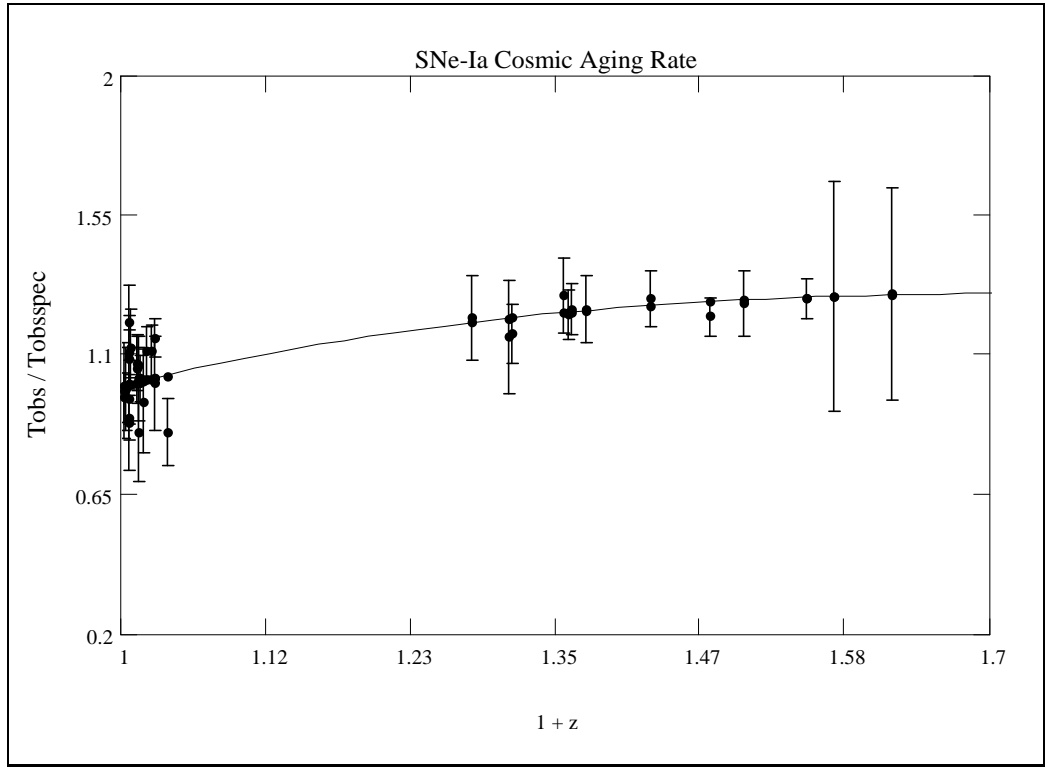}} 
\vspace*{8pt}
\caption{SCP low $z$ and high $z$ SNe-Ia light curve time intervals\cite[Blondin, et al.]{blondin-1}. 
Filled circles are $\Delta{t_{obs}} / \Delta{t'_{obsspec}}$ for each of the 35 SNe-Ia.  
Error bars are computed from the data errors. 
The solid line is the cosmic aging function $g_1(z)$. The reduced $\chi^2$ for the fit of $g_1(z)$ to the data is $\chi^2 / 34 = 0.690$. }
\label{fig:SCP-Blodin-1}
\end{figure}

\begin{figure}[htb!]
\fbox{\includegraphics[viewport=-5 295 430 585,keepaspectratio,clip=true]{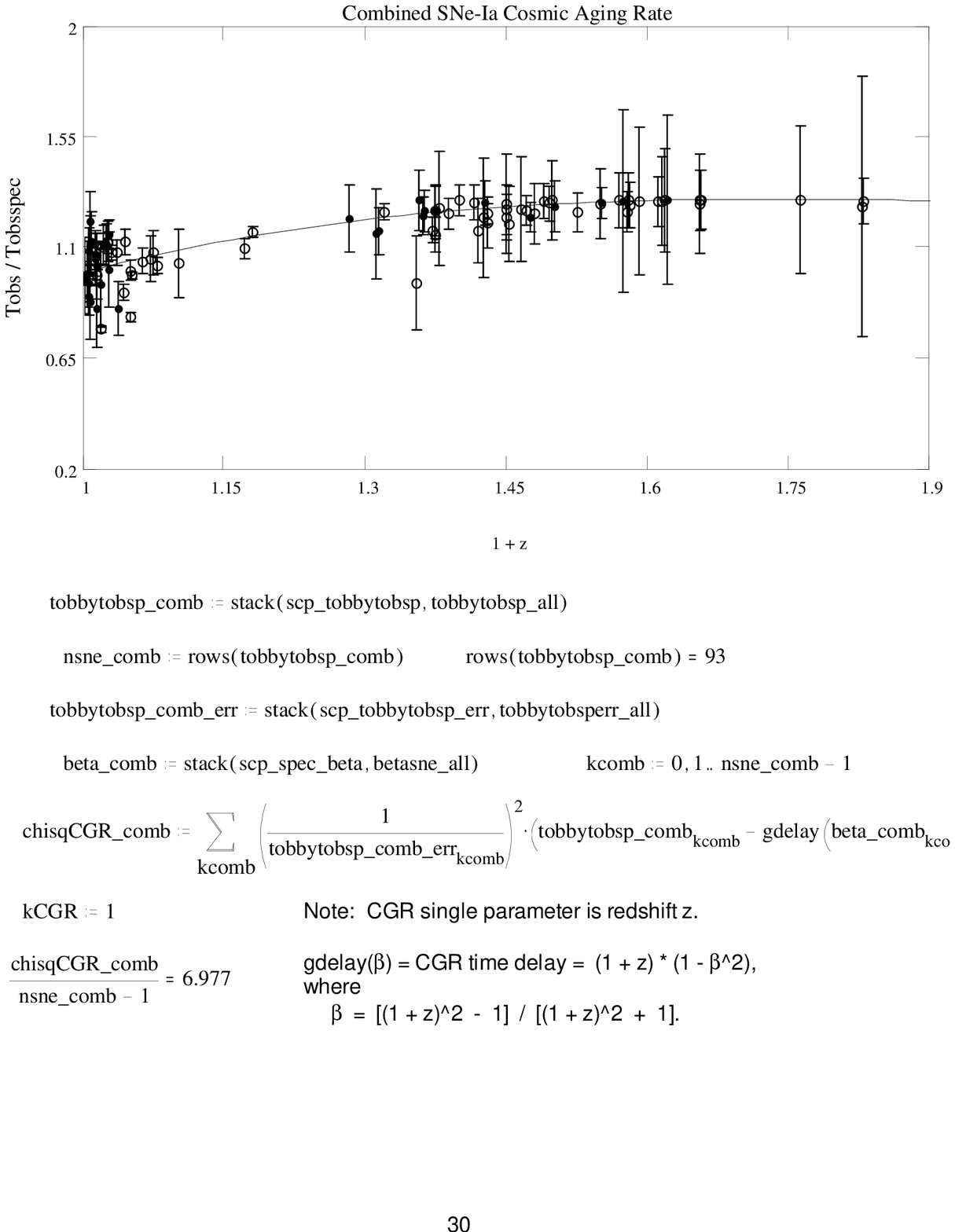}} 
\vspace*{8pt}
\caption{Combined SNe-Ia light curve time intervals $\Delta{t_{obs}} / \Delta{t'_{obsspec}}$.
Open circles are from\cite[Goldhaber, et al.]{goldhaber-1}.
Filled circles are from\cite[Blondin, et al.]{blondin-1}.
The solid line is the cosmic aging function $g_1(z)$. 
The reduced $\chi^2$ for the fit of $g_1(z)$ to the combined data is $\chi^2 / 92 = 6.98$. }
\label{fig:SCP-Combined-1}
\end{figure}

The next SNe data are from \cite{blondin-1}.  We take the aging rates $( \Delta{t_{spec}} /  \Delta{t_{obs}} )$
from \cite[Table 3]{blondin-1}, in the last column, unparenthesized.
We compute 
\begin{equation}
\frac{ \Delta{t_{obs}} }  { \Delta{t'_{obsspec}} } 
             = \left( \Delta{t_{obs}} / \Delta{t_{spec}} \right) \left( 1 - t^2 / \tau^2 \right). 
                    \label{eq:delta_tobs/delta_tobsspec-blondin}
\end{equation}
The errors come from \cite[Table 3]{blondin-1}, in the last column, parenthesized.  The redshifts  are computed
from the aging  rate data instead of from the given host galaxy values. The plotted data are shown in 
Fig. \ref{fig:SCP-Blodin-1}. Again we note  the reduction in the aging rate at higher redshifts. The reduced chi-squared for the
data fitted to the cosmic aging rate $g_1(z)$ is $\chi^2 / 34 = 0.690$. 
In Fig. \ref{fig:SCP-Combined-1} we show the combination of all the SNe aging data from the two reports. With 93 total
data points the reduced chi-squared is $\chi^2 / 92 = 6.98$.

\section{Simulation of quasar like light curve power spectra}

The purpose of this section is to simulate quasar like light curve power spectra to compare with the 
report of observed low and high redshift quasar light curve power spectra\cite[Hawkins]{hawkins-1}.
The observed light curve power spectra \cite[Fig. 5, left-hand panel]{hawkins-1}  were found to be identical
within the experimental errors.  Therefore, we will assume the low and high redshift light curves are identical 
in the observer frame $K$.  In addition, it is assumed that the redshifts are of pure cosmological origin, with no components
of gravitational redshifts or Doppler shifts.  In our simulation, the pseudo quasar light curve apparent magnitudes 
$m(j)$ at epoch $j$ is generated by the function
\begin{equation}
 m\left( j \right) =  {\rm exp}\left( -\frac{2 \pi j} {N_y } \right)  
                      {\rm  cos}^2\left( \frac{800 f_0 j} {N_e } \right)     
                      {\rm  sin}^2\left( \frac{200 f_0 j}  {N_e } \right),   \label{eq:Mqso}
\end{equation}
for each epoch $j= 1,2,.., N_e$, where $N_e = 560$, $N_y = 56$ years and 
$f_0 = 1 / 16.3 \, {\rm yr} = 0.0613 \, {\rm yr}^{-1}$. The redshifts used are low $z=0.765$ and high 
$z=1.711$ and  $f^{-1}_0 = 16.3 \, {\rm yr}$  are from the quasar time dilation report\cite{hawkins-1}. For better 
resolution we used $N_y = 56 \, {\rm yr}$ instead of the $28 \, {\rm yr}$ which was used  in the report.
The Fourier power spectrum $P_S(z, j)$ is determined from the magnitudes $m(j)$ by \cite[Equation (1)]{hawkins-1}
\begin{eqnarray}
P_S\left( z, j \right)  = & & \frac{ T\left( z \right) } { N_e }
         \left[ \sum_{k=1,N_e} { m\left( j \right) {\rm cos}\left( \frac{2 \pi j k} { N_e } \right) } \right]^2  \nonumber \\ 
\nonumber \\
        &+& \frac{ T\left( z \right)  } { N_e }
        \left[ \sum_{k=1,N_e} { m\left( j \right) {\rm sin}\left( \frac{2 \pi j k} { N_e } \right) } \right]^2,   \label{eq:Psj}
\end{eqnarray} 
where j runs over $N_e$ equally spaced epochs of simulated data separated by time $T\left( z \right)  / N_e$.
Then the time transformations will take us from the origin of observer frame $K$
to the quasar rest frame $K'$ at cosmic time $t$.  For CSR the sampling interval $T(z)$ we use is defined by
\begin{equation}
T\left( z \right)  =  \frac{ T_0 } { g_1\left( z \right) },  \label{eq:T(z)-csr}
\end{equation}
where $g_1(z)$ is given by (\ref{eq:g_1(z)-define}) and
\begin{equation}
T_0 = 1 / f_0.  \label{eq:T_0}
\end{equation}
We divide by $g_1(z)$ in (\ref{eq:T(z)-csr}) because we are going back in time to the quasar rest frame.

For our purposes, the flat space Friedmann-Lema\^{i}tre-Robertson-Walker (FLRW) model is the CSR model with the 
cosmic aging function $g_1(z)$ replaced by $(1 + z)$. For the flat space FLRW model we use for $T(z)$,
\begin{equation}
T\left( z \right)  = \frac{ T_0 } { 1 + z },  \label{eq:T(z)-flrw}
\end{equation}
where we have again divided out the time transformation to get to the quasar rest frame.

For either model we use the fitting function $P(f,z)$  defined by\cite[Equation (2)]{hawkins-1} 
\begin{equation}
P\left( f, z \right)  =  \frac{ C_0} {\left( f /  f_c\left( z \right)  \right)^a +  \left( f / f_c\left( z \right) \right)^{-b} },
      \label{eq:P-fit}
\end{equation}
where $C_0$ is the power,  $f$ is the frequency, $f_c(z)$ is the redshift dependent frequency at maximum power and $a$ and $b$
are constants.  We use the appropriate form of $f_c(z)$ for the CSR or the FLRW models. We show light curve power spectrum 
plots of ${\rm log}( P_S(z, j) \times f_j(z))$ vs. ${\rm log}(f_j(z))$ for the light curves at low and high redshift.

\begin{figure}[htb!]
\fbox{\includegraphics[viewport=-5 -5 225 280,keepaspectratio,clip=true]{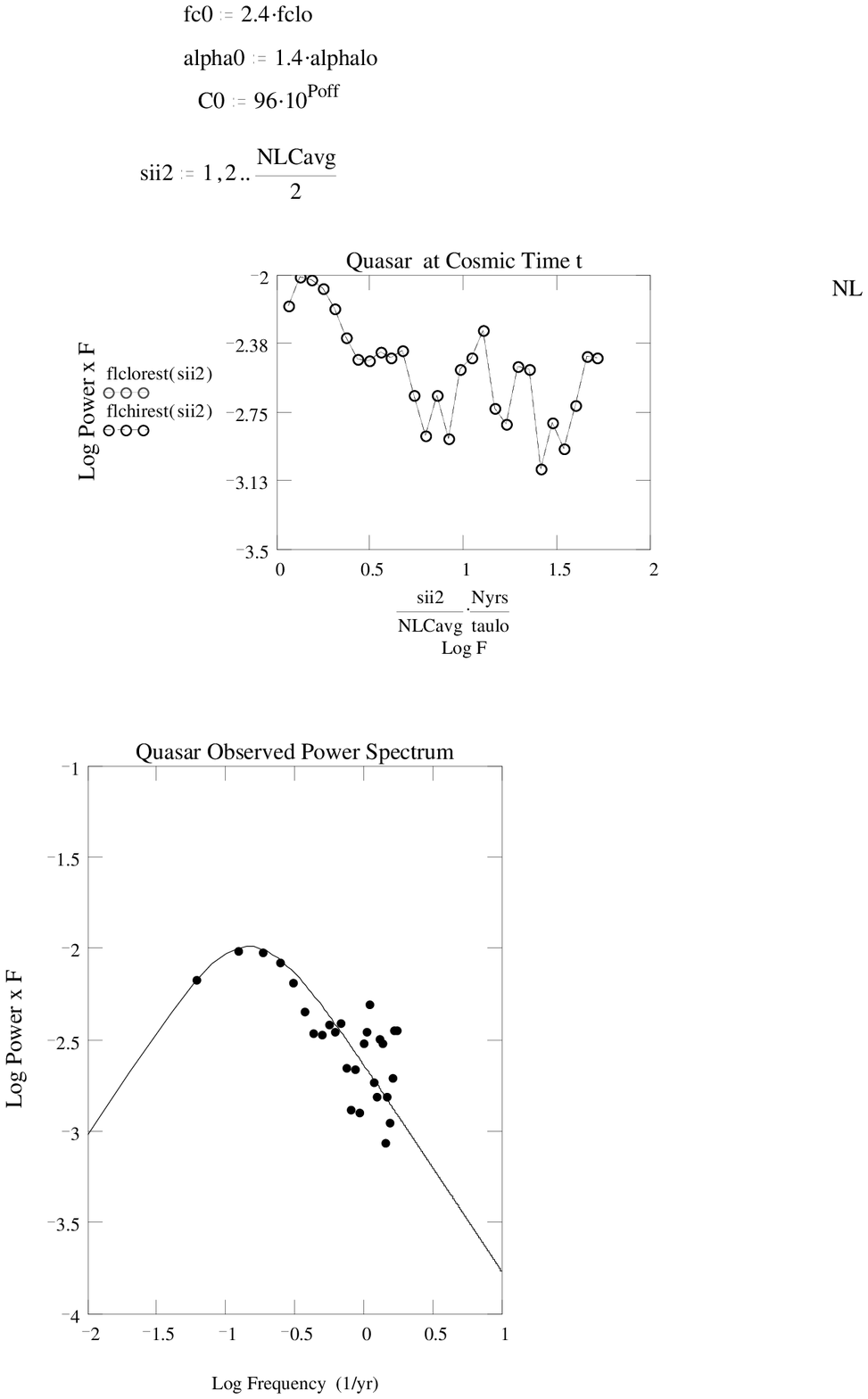}} 
\vspace*{8pt}
\caption{Simulated quasar light curve power spectrum, filled circles,  observed at origin of  frame $K$ at cosmic time $t=0$.
Low redshift and high redshift quasar power spectra overlap so are shown as one spectrum.
The frequency $f$ is ${yr}^{-1}$.
The  abscissa (horizontal) axis is ${\rm log}(f)$ and the ordinate (vertical) axis is ${\rm log}(power \times f)$.
Both axes have unit scaling.
The function $P_f(f)$, solid line, was  fitted iteratively with parameters  $C_0=0.020524$, 
$f_c(0)=0.14724 \, {\rm yr}^{-1}$, $a=1.134$ and $b=-a$, yielding $\chi^2 / N_y / 2 = 0.042673.$   }  
\label{fig:QSO-ObsPower-CSR-2}
\end{figure}

Assuming both quasars have identical power spectra in the observer frame $K$, we obtain the power spectrum from (\ref{eq:Psj})
by setting the redshift $z=0$ which is  given by $P_S(0,j)$.  This is shown in Fig. \ref{fig:QSO-ObsPower-CSR-2} with the fitting power 
function $P_f(f,z)$ parameters $C_0 = 0.020524$, $f_c(z=0) = 2.4 f_0$, $a = 1.4 \, \alpha$ where $\alpha = 0.81$ from 
\cite[(Table 1, Observer frame Sample= $z < 1$, Index=-0.81) ]{hawkins-1} and $b = a$.  This can be compared
with \cite[Fig. 5, left-hand panel.]{hawkins-1}

\begin{figure}[htb!]
\fbox{\includegraphics[viewport=-5 290 425 520,keepaspectratio,clip=true]{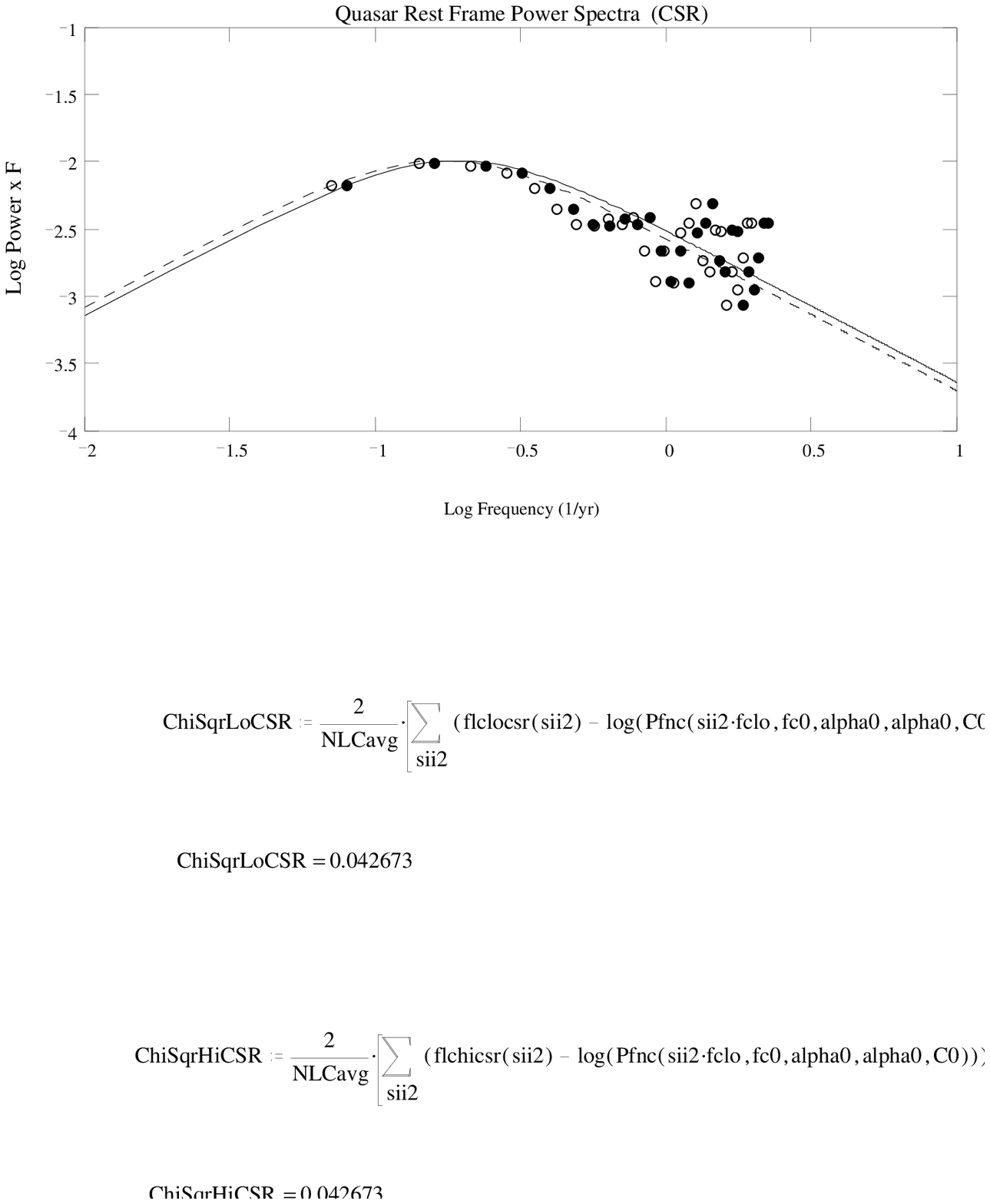}} 
\vspace*{8pt}
\caption{Simulated quasar light curve power spectra in the rest frame $K'$ of each quasar where each light curve transforms according 
to CSR.  The frequency $f$ is ${yr}^{-1}$.
For the low redshift power spectrum,  filled circles, $z_{low}=0.765$, the horizontal axis is scaled by $g_1(z_{low})$ 
with $g_1(z)$ from (\ref{eq:g_1(z)-define}) and the vertical axis has unit scaling. The power spectrum time 
interval is transformed by $1 / g_1(z_{low})$ and the frequency is scaled  by $g_1(z_{low})$ which effectively sets the scale to unity.
For the high redshift power spectrum, open circles, $z_{high}=1.711$, the horizontal axis  is scaled by $g_1(z_{high})$
and the vertical axis has unit scaling although the power spectrum time interval is transformed by $1 / g_1(z_{high})$ 
and the frequency is scaled by $g_1(z_{high})$.
The fitting function $P_f(f)$ has the same parameters  as used at the origin of $K$ except the central
frequency is scaled by $f_c(z)=0.14724 \times g_1(z)$ with $z=z_{low}$, solid line,  and  $z=z_{high}$, dashed line.  }
\label{fig:QSO-Power-Cosmic-t-2}
\end{figure}

Next we show the light curve power spectra for the low and high redshift quasars, $z_{low} = 0.765$ and $z_{high} = 1.711$, 
respectively, as it would be observed in their rest frame. 
The light curves are corrected by $1 / g_1(z)$ since we are obtaining the light curve back in time.
We show the quasar power spectrum  along with the fitting  function $P_f(f,z)$, which has the same parameters
as were used at the origin of $K$ except for the frequency at maximum power which is given by 
\begin{equation}
f_c(z) = 2.4 \, f_0  \, g_1(z),  \label{eq:f_c(z)-CSR}
\end{equation}
with $g_1(z)$ from (\ref{eq:g_1(z)-define}).  This is plotted in  Fig. \ref{fig:QSO-Power-Cosmic-t-2}. The fitting function 
has the same parameters for both the low and high redshift spectra as were used in the spectrum at the origin of $K$
except for $f_c(z)$.

\begin{figure}[htb!]
\fbox{\includegraphics[viewport=0 285 370 515,keepaspectratio,clip=true]{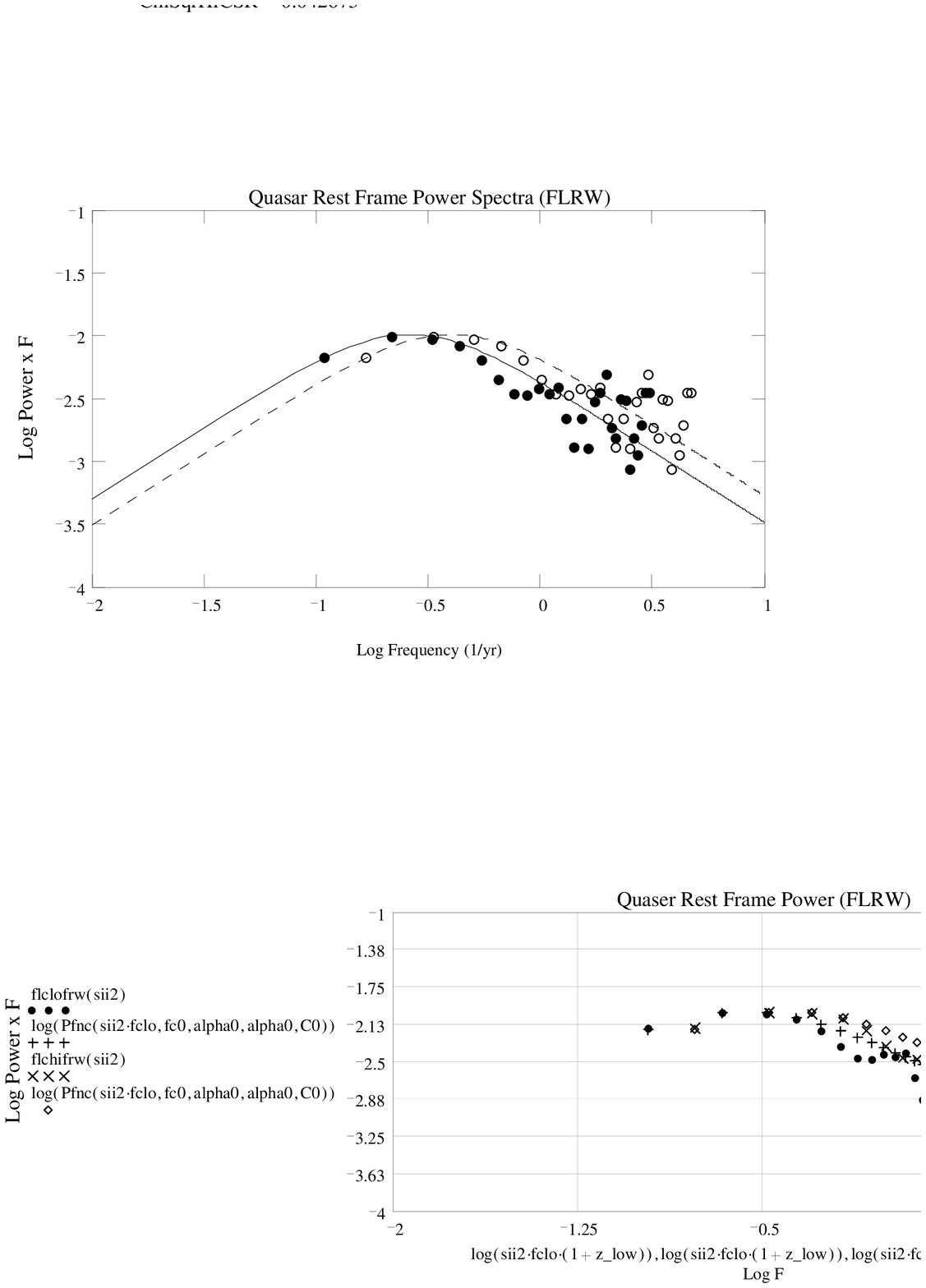}} 
\vspace*{8pt}
\caption{Simulated quasar light curve power spectra in the rest frame $K'$ of each quasar
where each light curve transforms according to flat space FLRW.
The frequency $f$ is ${yr}^{-1}$.
For the low redshift power spectrum,  filled circles, $z_{low}=0.765$, the horizontal axis is scaled by $1 + z_{low}$.
and the vertical axis has unit scaling. The power spectrum time 
interval is transformed by $1 / 1 + z_{low}$ and the frequency is scaled  by $1 + z_{low}$ which effectively sets the scale to unity.
For the high redshift power spectrum, open circles, $z_{high}=1.711$, the horizontal axis  is scaled by $1 + z_{high}$
and the vertical axis has unit scaling although the power spectrum time interval is transformed by $1 / 1 + z_{high}$ 
and the frequency is scaled by $1 + z_{high}$.
The fitting function $P_f(f)$ has the same parameters  as used at the origin of $K$ except the central
frequency is scaled by $f_c(z)=0.14724 \times (1 + z)$ with $z=z_{low}$, solid line,  and  $z=z_{high}$, dashed line.  }
\label{fig:QSO-Power-RestFrame-FLRW-2}
\end{figure}

To show what the power spectra might look like for a flat space FLRW observation we show the low and high redshift quasar
light curve power spectra when the light curves are corrected for time dilation by $1 / (1 + z)$, assuming a flat space cosmology. 
This is shown in Fig. \ref{fig:QSO-Power-RestFrame-FLRW-2}.  This plot is similar to \cite[Fig. 5, right-hand panel]{hawkins-1}.
The fitting function $P_f(f,z)$ has the same parameters for both the low and high redshift spectra as
were used at the origin of $K$ except for the frequency at maximum power which is given by 
\begin{equation}
f_c(z) = 2.4 \, f_0 \, (1+z).   \label{eq:f_c(z)-FLRW}
\end{equation}

\begin{figure}[htb!]
\fbox{\includegraphics[viewport=-10 -10 365 380,keepaspectratio,clip=true]{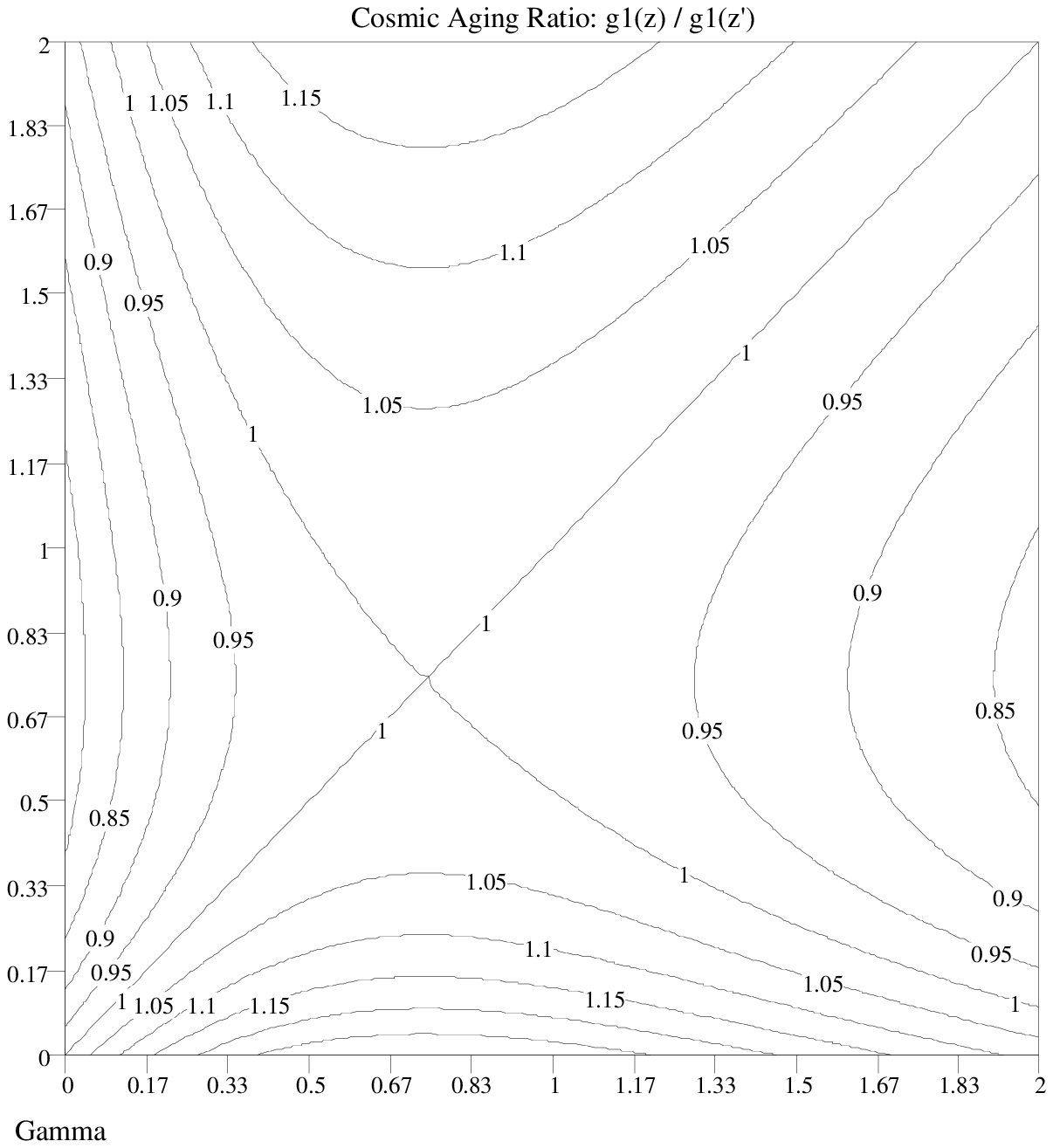}} 
\vspace*{8pt}
\caption{Contour plot  of ${\rm Gamma}(z,z') = g_1(z) / g_1(z')$ where 
 $g_1(z) =  4 ( 1 + z )^3 /  [ ( 1 + z )^2 + 1 ]^2$ from (\ref{eq:g_1(z)-define}).
The horizontal and vertical axes are redshift $z$. 
Of interest is the curved unity ($1$) contour going from horizontal lower right to vertical upper left in the graph. }
\label{fig:CSR-CosmicAgingRatio-1}
\end{figure}

In Fig. \ref{fig:CSR-CosmicAgingRatio-1} we show a contour plot of the cosmic aging ratio ${\rm \Gamma}\left( z,z' \right)$ 
defined by
\begin{equation}
{\rm \Gamma}\left( z,z' \right)  =  \frac{g_1\left( z \right) } {g_1\left( z' \right) },  \label{eq:Gamma(z,z')-def}
\end{equation}
between two source fields, one at redshift $z$ and the other at redshift $z'$.  This is to demonstrate that it is possible to
obtain  similar aging rates (eg. within $10 \, \%$) between two sources separated by large redshift.
For the above low and high redshifts, ${\rm \Gamma}(1.711, 0.765) = 0.8803$.

\begin{figure}[htb!]
\fbox{\includegraphics[viewport=-5 -10 430 230,keepaspectratio,clip=true]{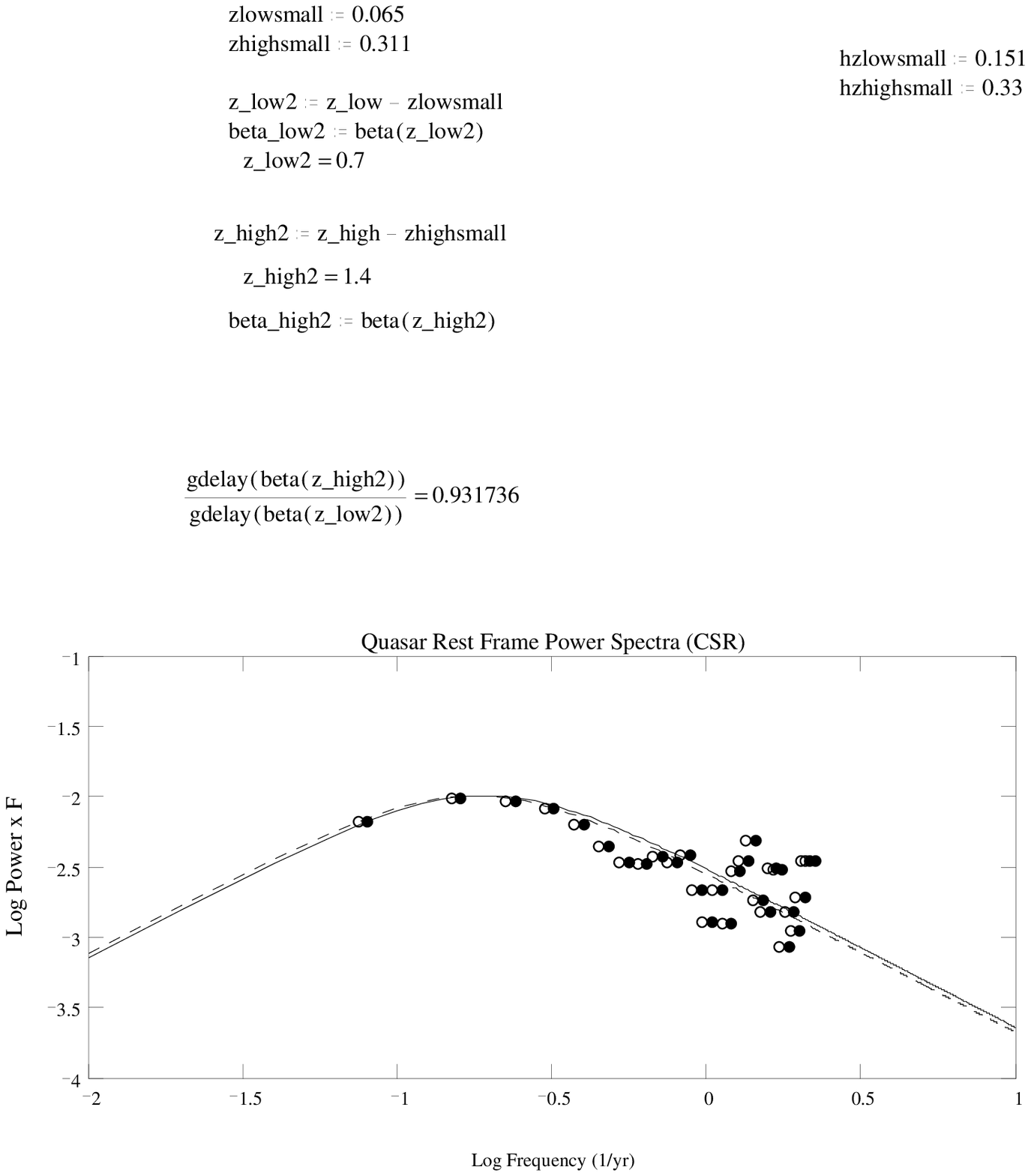}} 
\vspace*{8pt}
\caption{Simulated quasar light curve power spectra in the rest frame $K'$ of each quasar,
assuming each light curve transforms according to CSR.  
The frequency $f$ is ${yr}^{-1}$.
For the low redshift power spectrum,  filled circles, $z_1=0.7$, the horizontal axis is scaled by $g_1(z_1)$
and the vertical axis has unit scaling. The power spectrum time 
interval is transformed by $1 / g_1(z_1)$ and the frequency is transformed  by $g_1(z_1)$ which effectively makes
the scale unity.
For the high redshift power spectrum, open circles, $z_2=1.4$, the horizontal axis  is scaled by $g_1(z_2)$
and the vertical axis has unit scaling although the power spectrum time interval is transformed by $1 / g_1(z_2)$ 
and the frequency is scaled by $g_1(z_2)$.
The fitting function $P_f(f)$ has the same parameters  as used at the origin of $K$ except the central
frequency is scaled by $f_c(z)=0.14724 \times g_1(z)$ with $z=z_1$, solid line,  and  $z=z_2$, dashed line.  }
\label{fig:QSO-Power-Cosmic-t-2B}
\end{figure}

For another demonstration of the cosmic aging ratio, we show in Fig. \ref{fig:QSO-Power-Cosmic-t-2B} the power spectrum  
for each quasar in its rest frame where $z_1 = 0.7$ and $z_2 = 1.4$.  The aging ratio is ${\rm \Gamma}(z_2, z_1) = 0.9317$. 

\section{CMB anisotropy acoustic peak}

In CGR the big bang occurred at velocity $v=c$.  At the recombination of protons and electrons in the baryon-photon plasma,
when the photons decoupled to form the CMB radiation field, the velocity was $v=v_{rc}$, which is related\cite{carmeli-0}
to the time $t_{rc}$ of recombination by $v_{rc} / c  = t_{rc} / \tau$.  Applying (\ref{eq:t/tau_func(z)}), we have
\begin{equation}
 \frac{ v_{rc} } { c }  = \frac{ t_{rc} } { \tau }  = \frac{ \left( 1 + z_{rc} \right)^2  - 1 } { \left( 1 + z_{rc} \right)^2  + 1 },
       \label{eq:v_rc/c}
\end{equation}
where $z_{rc}$ is the cosmological redshift at recombination.  The coordinate distance $r_{bb}$ to the big bang is given by 
(\ref{eq:r-with-any-Omega}) with $v/c = 1$,
\begin{equation}
r_{bb}  =  \frac{ c \tau } { \sqrt{ 1 - \Omega_M} }  \, {\rm sinh}\left( \sqrt{ 1 - \Omega_M} \right).
                   \label{eq:r_bb} 
\end{equation}
Likewise, the coordinate distance $r_{rc}$ to the recombination epoch with $v/c = v_{rc} / c$,  is given by 
\begin{equation}
r_{rc}  =  \frac{ c \tau } { \sqrt{ 1 - \Omega_M} }  \, {\rm sinh}\left( \frac{v_{rc}} {c} \sqrt{ 1 - \Omega_M} \right).
                   \label{eq:r_rc} 
\end{equation}
We construct a simple model to determine the size of the sound horizon\cite{cornish-1,frieman-1} for the longest sound wave,
which generates the first acoustic peak.  If $r_e$ is the radius of the sphere of expanding plasma,
$\bar{v}_e = (c + v_{rc})/2$ is the average expansion velocity between the big bang and the recombination epoch
and $c_s = c / \sqrt{3}$ is the speed of the longest sound wave in the plasma,  then, by proportion of velocities,
$r_{sh} = (c_s / \bar{v}_e ) r_e$ is the radius of the sphere containing the longest wave.
Assuming that the wave travels along a great circle path of the sphere, the size of the sound horizon $d_{sh}$
is given by
\begin{equation}
d_{sh}  = 2 \pi r_{sh}  = \frac{ 4 \pi r_e} { \sqrt{3} \left( 1 + v_{rc}/c \right) }.  \label{eq:d_sh-model}
\end{equation}
Defining $r_e = r_{bb} - r_{rc}$, which is  the difference of (\ref{eq:r_bb}) and (\ref{eq:r_rc}), then from 
(\ref{eq:d_sh-model}), the size of the sound horizon at recombination is given by 
\begin{eqnarray}
d_{sh} &=& \frac{4 \pi } {\sqrt{3} } \frac{ \left(  r_{bb} - r_{rc} \right) } { \left( 1 + v_{rc} / c \right) }  =  \label{eq:d_rc} \\
\nonumber \\
          &=&  \frac{4 \pi c \tau } { \sqrt{3} } 
                    \left\{ \frac{ {\rm sinh}\left[ \sqrt{ 1 - \Omega_M} \right]
                                    -  {\rm sinh}\left[ \left( v_{rc} / c \right)  \sqrt{ 1 - \Omega_M} \right]  } 
                                        {  \sqrt{ 1 - \Omega_M} \left(1 + v_{rc} / c \right) } 
                    \right\}.  \nonumber
\end{eqnarray}
The angle $\theta_{sh}$ of the sound horizon at recombination is given by
\begin{equation}
\theta_{sh}  = \frac{d_{sh}} {DA_{rc}},   \label{eq:theta_sh_def}
\end{equation}
where the angular diameter distance $DA_{rc}$ is given by\cite{hartnett-1}
\begin{equation}
DA_{rc}  = \frac{DL_{rc}} {\left( 1 + z_{rc} \right)^2 },   \label{eq:DA_rc}
\end{equation}
where $DL_{rc}$ is the luminosity distance of the recombination epoch. Substituting from (\ref{eq:v_rc/c})-(\ref{eq:d_rc})
 and (\ref{eq:DA_rc})
into (\ref{eq:theta_sh_def}) and simplifying we obtain
\begin{equation}
\theta_{sh}  =  \frac{ 4 \pi } {\sqrt{3}} \left[ \frac{ \left( 1 + z_{rc} \right) \sqrt{ 1 - t^2_{rc} / \tau^2 } }
                                                                                 { 1 +  v_{rc} / c } \right]
    \left\{ \frac{ {\rm sinh}\left[ \sqrt{1 - \Omega_M } \right] } 
                   { {\rm sinh}\left[ \left(  v_{rc} / c \right) \sqrt{1 - \Omega_M } \right]  }
             - 1 \right\}.   \label{eq:theta_sh_expand}
\end{equation}
The CMB radiation escaped the matter sphere and expanded to fill all space. 
The size of the sound horizon $d_{sh0}$ in the CMB on today's sky is obtained by applying 
(\ref{eq:cosmo-redshift}) with $d_{sh}$,
\begin{equation}
d_{sh0}  = d_{sh} \left( 1 + z_{rc} \right).  \label{eq:X_sh}
\end{equation}
Then, the angle $\theta_{sh0}$ of the sound horizon in the CMB radiation field on today's sky is given by 
\begin{equation}
\theta_{sh0}  = \frac{d_{sh0}} {DA_{rc}} = \theta_{sh} \left( 1 + z_{rc} \right).  \label{eq:theta_sh_today}   
\end{equation}
The multipole $l$ of the first acoustic peak\cite{cornish-1} recorded in the CMB radiation field is proportional 
to the inverse of (\ref{eq:theta_sh_today}),
\begin{equation}
l \approx \frac{\pi} {\theta_{sh0}} = \frac{\pi} {\theta_{sh} \left( 1 + z_{rc} \right) }.  \label{eq:l_1}
\end{equation}
Substituting  $\Omega_M = 0.800 \pm 0.080$ and $z_{rc} = 1100$ into the above equations we obtain a value
of 
\begin{equation}
l \approx 224 \pm 5  \label{eq:ell_1},
\end{equation}
which is in good agreement with observation\cite{hinshaw-1,hanany-1}
and,
\begin{equation}
\theta_{sh0} \approx  0.805^{\circ} \pm 0.020^{\circ}.  \label{eq:theta_0_degree} 
\end{equation}
The size of the sound horizon on today's sky is $d_{sh0} \approx 30.2 \, {\rm Mpc}$, 
which is $1/5$ the value of the standard model.

\section{Discussion}

Let us review briefly some aspects of  the Carmeli five dimensional brane world cosmological model.  

\subsection{ Velocity, acceleration and cosmic distances in CSR  }

From (\ref{eq:euclidean-metric}), for $dt = 0$ with $dx^2 + dy^2 + dz^2 = dr^2$
we have,
\begin{equation}
  ds^2 = \tau^2 dv^2 - \left( dx^2 + dy^2 + dz^2 \right),  \label{eq:flat-space-metric}
\end{equation}
This can be manipulated to obtain
\begin{equation}
1 = \tau^2 \left( \frac{ dv } { ds } \right)^2 \left(  1 - \frac{dx^2 + dy^2 + dz^2 } { \tau^2 dv^2}  \right) 
   = \tau^2 \left( \frac{ dv } { ds } \right)^2 \left( 1 - \frac{ t^2 } { \tau^2 } \right),   \label{eq:tau dv/ds=1}
\end{equation}
where 
\begin{equation}
t^2 = \left(dx^2 + dy^2 + dz^2 \right) / dv^2  \label{dr/dv=t}
\end{equation}
is the cosmic time (squared).   Using (\ref{eq:tau dv/ds=1}) this gives for the components of the four-velocity in CSR,
\begin{equation}
u^{\mu} = \frac{dx^{\mu}} {ds} = \frac{dx^{\mu}} {dv} \frac{dv} {ds}
                = \frac{1} {\tau \sqrt{ 1 - t^2/ \tau^2 } } \frac{ dx^{\mu}} { dv }
                = \frac{ \gamma } { \tau } \frac{ dx^{\mu}} { dv },  \label{eq:4-velocity}
\end{equation}
where $\mu = 0, 1, 2, 3$ and
\begin{equation}
\gamma = \frac{1} {\sqrt{ 1 - t^2/ \tau^2 } }.  \label{eq:gamma-def}
\end{equation}
We have from (\ref{eq:4-velocity}) that
\begin{equation}
u^0 = \gamma, \label{eq:u0-def}
\end{equation}
\begin{equation}
u^k = \frac{\gamma} {\tau} \frac{dx^k} {dv},  \, \, k = \{ 1, 2,  3 \}. \label{eq:dxk-def}
\end{equation}
Defining $u_{\mu} = u^{\mu}$ we obtain for the invariant 4-vector length, from (\ref{eq:4-velocity}),
(\ref{eq:u0-def}) and (\ref{eq:dxk-def}),
\begin{equation}
u_{\mu} u^{\mu} = u_0 u^0 - u_1 u^1 - u_2 u^2 - u_3 u^3 = 1,  \label{eq:u-length}
\end{equation}
that is, the length of $u^{\mu}$ is unity in all CSR frames of reference.
Multiplying (\ref{eq:tau dv/ds=1}) by $a_0 \tau^4$, where $a_0$ is the ordinary acceleration measured in
the cosmic frame at time $0$, the local frame, we obtain after some manipulation,
\begin{equation}
a^2 \tau^4  - a^2 t^2 \tau^2  = a^2_0 \tau^4,  \label{eq:acceleration-a}
\end{equation}
where
\begin{equation}
a = \frac{a_0} {\sqrt{1 - t^2 /\tau^2 } },  \label{acceleration-at-cosmic-time}
\end{equation}
is the acceleration at any cosmic time $t$.   Equation (\ref{eq:acceleration-a}) can be put
into the form
\begin{equation}
a^2 \tau^4 - \tau^2 v^2 = a^2_0 \tau^4,  \label{eq:acceleration-and-velocity}
\end{equation}
where
\begin{equation}
v = a t = \frac{ a_0 t } { \sqrt{1 - t^2 /\tau^2 } } \label{eq:v=at}
\end{equation}
is the velocity of a point which had an acceleration $a$ over a time $t$.  Defining the cosmic distance $S=a \tau^2$
we have from (\ref{eq:acceleration-and-velocity})
\begin{equation}
S^2 - \tau^2 v^2  = S^2_0, \label{S2} 
\end{equation}
where $S_0 = a_0 \tau^2$.  This is analogous to the energy equation in SR, $E^2 - c^2 p^2 = E^2_0$.
Refer to \cite{carmeli-4} for a thorough treatment of this topic.

\subsection{ Behavior for large cosmic time  }

The cosmological redshift (\ref{eq:cosmo-redshift-final}) and  the cosmic aging function (\ref{eq:g_1(z)-define}) 
are two functions which can be used to describe the behaviour expected at large cosmic time $t \approx \tau$, 
where $\tau$ is the Hubble-Carmeli time constant and is the largest possible time.   For the cosmological redshift,
for observations of events close to the big bang we have
\begin{equation}
 1 + z  \propto \frac{ 1 } {\sqrt{ 1 - t / \tau } }  \rightarrow \infty \, \, {\rm as} \, \,  t \rightarrow \tau. 
   \label{eq:(1+z)limit}
\end{equation}
The luminosity distance (\ref{eq:Lum-dist-tbytau}), as $t \rightarrow \tau$,  $v/c \rightarrow 1$,
has the form
\begin{equation}
D_L\left( t \right) \propto \frac{ c \tau  }  { \left( 1 - t / \tau \right)  } \rightarrow \infty
               \, \, {\rm as} \, \,  t \rightarrow \tau.
            \label{eq:DL(t->tau)}
\end{equation}
For the standard model, the luminosity distance relation, by (\ref{eq:(1+z)limit}), 
$DL_{SM} \propto c \tau / \sqrt{1-t/\tau}$, for large $t$.
We see that the CGR luminosity distance, by (\ref{eq:DL(t->tau)}), is larger than the standard model
by the factor $1/\sqrt{1-t/\tau}$.
On the other hand, from the cosmic aging function (\ref{eq:f(t/tau)}), for an observation $\Delta{t}$ of an 
elapsed time $\Delta{t'}$ which occurred close to the big bang time we have
\begin{equation}
\Delta{t}  = \Delta{t'} g_1(t) =    \Delta{t'} \left( 1 + t / \tau \right)  \sqrt{ 1 - t^2 / \tau^2} 
                \rightarrow 0  \, \, {\rm as} \, \,  t \rightarrow \tau. 
\end{equation}
This implies that durations of events, such as for example star formation, star collapse or star bursts, 
observed in nearby galaxies at cosmic times $t / \tau < 0.8395$ (redshifts $z < 3.385$) should be observed to have
shorter durations the further back we look beyond cosmic times $t/ \tau > 0.8395$ (redshifts $z > 3.385$.)

\subsection{The cosmological redshift vs. the cosmic aging function}

One may ask why the cosmological redshift of  the wave length ${\lambda}$ of light is given by the relation
$(1 + z) {\lambda}$ 	instead of with the cosmological aging function $g_1(t) {\lambda}$ . The answer is that
light wave phenomena do not involve the addition of cosmic times as do evolutionary phenomena such as 
a star burst or collapse. Light propagation is only affected by cosmic expansion while evolutionary phenomena
are affected by cosmic time addition and cosmic expansion.

\subsection{The accelerated expansion}

CGR does not have a cosmological constant, but it does have a critical mass density $\rho_c$.  From 
(\ref{eq:rho_eff-def}),  the effective mass density can be defined in terms of a vacuum mass density
\begin{equation}
\rho_{eff} = \rho + \rho_{vac},  \label{eq:rho_eff_vacuum}
\end{equation}
where
\begin{equation}
\rho_{vac} = -\rho_c = -3 / 8 \pi G \tau^2  \label{eq:rho_vac-def}
\end{equation}
is the constant {\em negative} mass density of the vacuum, which is not the common view of a vacuum density.
Differentiating  (\ref{eq:dr-by-dv-solve}) with respect to $v$ we obtain the acceleration in space-velocity, which
we put in the form
\begin{equation}
\frac{d^2 r } { dv^2 } + K r  = 0,  \label{eq:hooks-law-of-universe}
\end{equation}
where 
\begin{equation}
K =  \left( \rho + \rho_{vac} \right) /  c^2 \rho_c  = \left( \Omega - 1 \right) / c^2,  \label{eq:k-def}  
\end{equation}
and we have made the substitution, using (\ref{eq:rho_vac-def}),
\begin{equation}
\Omega_{vac} =  \rho_{vac} / \rho_c = -1.  \label{eq:Omega_vac-def}
\end{equation}
Equation (\ref{eq:hooks-law-of-universe}) is Hooke's law of the universe\cite[Section 5.4]{carmeli-2}
where $K$ is Hooke's constant for the universe.  If $\Omega > 1$ then $K$ is positive and its solution is 
a sum of sine and cosine functions and the universe has a decelerated expansion and is closed. 
If $\Omega < 1$ then $K$ is negative and the solution is a sum of hyperbolic  sinh and cosh functions, which means
the universe has an accelerated expansion and is open; this is the situation in our universe today where
we derived $\Omega = \Omega_M = 0.800$.  If $\Omega = 1$ then $K=0$ and the universe is not accelerating
and is neither open nor close. 

Although beyond the scope of this paper, we give an expression\cite{oliveira-1} for the vacuum density $\rho_{vac}$ in 
relation to the Bekenstein-Hawking black hole entropy\cite{bh-entropy} given by $S = ( k c^3 A ) / ( 4 \hbar G )$,
 where $k$ is Boltzmann's constant, $\hbar$ is Planck's constant over $2 \pi$ and  $A = 4 \pi c^2 \tau^2$ is the area 
 of the event horizon. For our universe  of mass $M = \tau c^3 / 4 G$,
 where the universe radius is twice the Schwarzschild radius,  the entropy is given by
  \begin{equation}
           S = \frac{\pi  k \tau^2 c^5 } { \hbar G },
 \end{equation}
 which can be put into the form relating to the vacuum mass density
 \begin{equation}
      \rho_{vac} =  \frac{-3}{8 \pi G \tau^2} = \frac{\rho_P } { \left( S / k \right) },
 \end{equation}
  where  the cosmological Planck mass density $\rho_P = -{\cal M}_P /  L^3_ P$. The  cosmological  Planck mass
   ${\cal M}_P = \sqrt{ \sqrt{ 3 / 8 } \, \hbar c / G}$  and  length $L_P = \hbar /   {\cal M}_P c$.
   The value of $ (S/k) \approx 1.980 \times 10^{122}$.

\subsection{Gravitational waves as a theoretic selection criteria }

When gravitational waves are detected we will be able to better quantify 
the strengths and weaknesses of the standard model (GR) and other models.  In this regard,  a paper on 
gravitational wave interferometry\cite{corda-1} has a good description of alternative theories to GR and is a 
fine starting place for further research.  Along those lines of inquiry, \cite{hartnett-2} describes the behaviour
of gravitational waves in Carmeli cosmology,  predicting a highly attenuated result for gravitational 
waves from galactic sources but possible detectability for gravitational waves from within the Milky Way Galaxy.

\section{Conclusion}

In this paper we used the linearized approximation of the 5-D Cosmological General Relativity as developed by 
Carmeli. A flat space CSR model was derived in a general way from the curved space CGR model.  
The CGR luminosity distance relation was applied to SCP Union2.1 SNe-Ia distance data up to redshift $z < 1.5$
combined with GRB distance data up to redshift $z < 7$. Utilizing the reduced 
$\chi^2$ method in the data analysis it is shown that the CGR model with a best fit mass density of  
$\Omega_M = 0.800 \pm 0.080$  performed as well as the ${\Lambda}CDM$ flat space model with an apriori 
mass density of $\Omega_M = 0.270$.  Regarding the hypothetical X particle constituents, the CGR model determines
rest mass energies for the $\bar{Y}$ and $\Phi^{*}$ particles of 
$m_{\bar{Y}} c^2 = m_{\Phi^{*}} c^2  \approx  9.79  \pm  0.47 \, {\rm GeV}$.  
We also found that CSR can confirm the $\Lambda{\rm CDM}$ model result of time dilation of $(1 + z)$ in SNe-Ia light 
curves. We also showed how the cosmic aging function $g_1(z)$ produces a null effect of time dilation in simulated 
light curve power spectra between two groups of hypothetical QSO's separated by a redshift $\Delta{z} \approx 1.0$.
Finally, we gave a model for obtaining the first acoustic peak of the CMB anisotropy, deriving a multipole 
$l \approx 224 \pm 5$, in good agreement with observation, and with an angle on the sky of 
$\theta_{sh0} \approx  0.805^{\circ} \pm 0.020^{\circ}$.

\begin{appendix}

\section{Fixed acceleration $dv/dt = c / \tau$  \label{ap:dv/dt} }

We show that $dv/dt = c / \tau$.  From (\cite{carmeli-2}, Sect. 7.4.2, (7.4.16a) and (7.4.16b)),
\begin{eqnarray}
\frac{d^2 x^k} {dt^2}  &=&  G M \frac{\partial} {\partial{x^k}}  \left( \frac{1} {r} \right), \label{eq:d2x/dt2} \\
\nonumber \\
\frac{d^2 x^k} {dv^2}  &=&  \frac{\tau^2 G M} {c^2} \frac{\partial} {\partial{x^k}}  \left( \frac{1} {r} \right),
       \label{eq:d2x/dv2} 
\end{eqnarray}
for $k = 1, 2, 3$.  Taking the left hand side of (\ref{eq:d2x/dv2}), 
\begin{eqnarray}
\frac{d^2 x^k} {dv^2}  &=&  \frac{d} {dv} \left( \frac{d x^k} {dv} \right),   \label{eq:d2x/dv2-A}  \\
\nonumber \\
                                  &=&  \frac{d} {dt} \left(  \frac{d x^k} {dt} \frac{dt} {dv} \right) \frac{dt} {dv}, 
                                       \label{eq:d2x/dv2-B}  \\
\nonumber \\
                                   &=&  \left[ \left( \frac{d^2  x^k } {dt^2} \right) \frac{dt} {dv} 
                                         + \frac{d  x^k } {dt} \frac{d} {dt} \left( \frac{dt} {dv} \right) \right] \frac{dt} {dv}.  
                                        \label{eq:d2x/dv2-C} 
\end{eqnarray}
The second term in (\ref{eq:d2x/dv2-C}) vanishes because
\begin{equation}
\frac{d} {dt} \left( \frac{dt} {dv} \right)  = \frac{d} {dv} \left( \frac{dt} {dt} \right) = \frac{d} {dv} \left( 1 \right)  = 0.
         \label{eq:d2t/dtdv}
\end{equation}
This then leaves
\begin{equation}
\frac{d^2 x^k} {dv^2}  =   \frac{d^2  x^k } {dt^2}  \left( \frac{dt} {dv} \right)^2.  \label{eq:d2x/dv2-D}
\end{equation}
Combining (\ref{eq:d2x/dt2}), (\ref{eq:d2x/dv2}) and (\ref{eq:d2x/dv2-D}) and simplifying leads to our intended result,
\begin{equation}
\frac{dv} {dt}  = \pm  \frac{c} {\tau}.  \label{eq:dv/dt=c/tau}
\end{equation}
Integrating the positive result of (\ref{eq:dv/dt=c/tau}), with 0 initial constants, yields
\begin{equation}
v = \frac{c} {\tau} \, t,   \label{eq:v=(c/tau)*t}
\end{equation}
which can be restated as
\begin{equation}
\frac{v} {c}  = \frac{t} {\tau}.  \label{eq:v/c=t/tau}  
\end{equation}

\end{appendix}

\end{document}